\documentclass[twocolumn]{aastex7}

\usepackage{amsmath}
\defcitealias{Dewberry2024}{D24}
\allowdisplaybreaks

\begin{document}

\title{Tidally driven inertial waves enhance eccentricity damping and spin evolution in planets and stars}

\author[0000-0001-9420-5194]{Janosz W. Dewberry}
\affiliation{Department of Astronomy, University of Massachusetts Amherst, 710 N Pleasant St, Amherst, MA 01003, USA}
\email[show]{jdewberry@umass.edu}  

\begin{abstract}
Tidal interactions influence the orbital motions of binary star systems and extrasolar planets alike. Tides also affect stellar and planetary rotation rates. We demonstrate that in addition to altering spin synchronization and pseudosynchronization, tidally driven inertial waves in the convective envelopes of low-mass stars and gas giant planets can enhance tidal eccentricity damping. Analytically, we find that eccentricity damping by inertial waves can be orders of magnitude faster than equilibrium tides, independent of any eddy viscosity prescription. We use simplified numerical experiments to demonstrate this enhancement, and to explore the effects of different mixing length treatments of convective turbulence, as well as a spin-down torque from magnetic braking. These calculations demonstrate that tidally driven inertial waves can produce an extended cool core of nearly circular binaries, helping to reconcile a longstanding discrepancy between observed and predicted main-sequence binary circularization. Our calculations additionally suggest that tidally driven inertial waves may leave identifiable signatures in the ratios of orbital to rotation periods for stellar binaries, including synchronous and sub-synchronous rotation periods reminiscent of populations identified in \emph{Kepler}, \emph{TESS}, and \emph{Gaia} data.
\end{abstract}

\vspace{-1em}


\section{Introduction}
The eccentricities of binary star systems provide compelling evidence that tidal interactions influence the orbital evolution of stellar and planetary systems. Tidal eccentricity damping is the generally accepted explanation for the observation that binaries comprising main-sequence stars exhibit predominantly circular orbits out to orbital periods as large as $\sim10-15$ days \citep{Meibom2005,Meibom2006,Mazeh2008}. Tides raised in evolved, giant stars appear to circularize much wider orbits \citep{Verbunt1995,Price-Whelan2018}. 

``Equilibrium'' tidal models in which this eccentricity damping results from the interaction of convective turbulence with nearly hydrostatic tidal bulges provide a reasonably effective explanation for the latter case \citep[with some caveats;][]{Beck2018,Benbakoura2021,Dewberry2025}. However, equilibrium tides fail to explain the circularization of main-sequence, solar-type binaries \citep[e.g.,][]{Terquem1998}, although the extent of this failure depends on contentious prescriptions for the eddy viscosity commonly used to describe the effect of convective turbulence on the tidal flow \citep{Zahn1989,Goldreich1977,Goodman1997,Penev2009,Vidal2020,Duguid2020,Terquem2021,Barker2021,Terquem2023}. While non-hydrostatic, ``dynamical'' tidal driving of internal gravity waves in the radiative cores of these stars enhances predicted rates of eccentricity damping \citep{Goldreich1989,Goodman1998,Terquem1998,Ogilvie2007,Barker2010,Zanazzi2021,Guo2023}, gravity waves still struggle to explain observed solar-type circularization \citep{Goodman1998,Barker2022}.

The rotation rates of tidally interacting planets and stars provide additional insight into their internal flows. Basic arguments assuming that angular momentum is conserved while orbital energy is dissipated lead to the prediction that tides will eventually drive rotation periods toward synchronization with the orbital period in circularized binaries \citep[if the total angular momentum of the system is large enough; e.g.,][]{Ogilvie2014}. Indeed, binaries detected by \emph{Kepler}, \emph{TESS}, and \emph{Gaia} exhibit synchronized surface rotation rates within $P_{\rm orb}\lesssim10$ days \citep{Lurie2017,Santos2019,Santos2021,Yu2025,Hobson-Ritz2025}. 

However, theoretical predictions are more complicated for spin evolution in binaries that are still eccentric. Simple constant time lag models for tidal energy dissipation predict evolution toward ``pseudosynchronous'' states in which rotation rates roughly match the instantaneous orbital frequency at periapsis \citep{Hut1981,Leconte2010}. But the assumption of a constant time lag is inappropriate even for equilibrium tides, unless convection somehow acts as an entirely frequency-independent effective viscosity \citep{Ivanov2004}. Moreover, constant time lag models ignore tidal driving of inertial waves that are intrinsic to the convective regions of rotating stars \citep[for example, inertial oscillation modes have now been detected in the Sun;][]{Loptien2018,Gizon2021,Hanson2022,Triana2022}. \citet{Dewberry2024} demonstrated that inertial waves can qualitatively alter tidal spin evolution in binaries on (artificially) fixed eccentric orbits, driving rotation rates toward spin states in which near-resonant driving of an inertial mode balances the equilibrium tidal torque. 

Tidal eccentricity damping and spin synchronization are closely linked. This paper demonstrates that the inertial wave-modified spin evolution highlighted by \citet{Dewberry2024} can in turn amplify tidal eccentricity damping, in some cases by orders of magnitude. The detailed, inherently tidal frequency-dependent mechanism considered here expands on the model of \citet{Barker2022}. The latter author used frequency-averaged calculations to argue that the tidal excitation of inertial waves in eccentric binaries is sufficient to reconcile theory with observations of main-sequence binary circularization. We similarly find tidally driven inertial waves extend circularization to larger orbital periods, although---in the absence of the impulsive approximation made by \citet{Barker2022}---reconciliation with observations depends on the degree to which convective turbulence acts as an eddy viscosity. Regardless, we find that the spin evolution intrinsic to our mechanism of eccentricity damping produces identifiable signatures in the ratios of orbital to rotation periods. 

This paper is structured as follows. Section \ref{sec:mech} derives an analytical estimate of the degree to which tidally driven inertial waves can amplify eccentricity damping over equilibrium tides. Section \ref{sec:model} then describes the interior and tidal models used to conduct numerical experiments described in Section \ref{sec:res}. Section \ref{sec:conc} presents discussion and conclusions.

\section{Mechanism}\label{sec:mech}
\subsection{Evolutionary equations}
Consider a primary object of mass $M$, equatorial radius $R$, and spin angular momentum $J$, in a coplanar orbit with a secondary object with mass $M'=qM$. On secular timescales, the orbit and the primary's angular momentum evolve according to 
\begin{align}
\label{eq:adot}
    \frac{\dot{a}}{a}
    &=\frac{\mathcal{P}}{E_o},
\\
\label{eq:edot}
    \frac{\dot{e}}{e}
    &=\frac{(1-e^2)}{e^2}
    \left(
        \frac{1}{2}\frac{\mathcal{P}}{E_o}
        +\frac{\mathcal{T}}{L_o}
    \right),
\\
\label{eq:Jdot}
    \dot{J}
    &=\mathcal{T}.
\end{align}
Here $a$ is the orbital semimajor axis, $e$ is the orbital eccentricity, $E_o=-GMM'/(2a)$ is the orbital energy, 
$L_o=-2(1 - e^2)^{1/2}E_o/n$ is the orbital angular momentum, and $n=[G(M+M')/a^3]^{1/2}$ is the mean motion. Lastly, $\mathcal{P}=-\dot{E}_o$ and $\mathcal{T}=-\dot{L}_o$ are time-averaged rates of energy and angular momentum transfer in a non-rotating frame with the origin fixed at the primary's center of mass. 

If this transfer is due solely to tides, and the tidal potential is truncated at quadrupolar order, the power $\mathcal{P}$ and torque $\mathcal{T}$ can be written as
\begin{align}
\label{eq:P}
    \mathcal{P}
    &=-\frac{1}{2}E_o\epsilon_Tn
    \sum_{k=-\infty}^\infty 
    k\left(
        |X_{0k}|^2\kappa_{0k}
        +3|X_{2k}|^2\kappa_{2k}
    \right),
\\\label{eq:T}
    \mathcal{T}
    &=-3E_o\epsilon_T
    \sum_{k=-\infty}^\infty |X_{2k}|^2\kappa_{2k}.
\end{align}
Here $\epsilon_T=q(R/a)^5,$ and $X_{mk}=X_{mk}(e)$ are Hansen coefficients \citep{Hughes1981} associated with a Fourier decomposition of the tidal potential into harmonic driving terms $\propto Y_{2m}(r/R)^2\exp[-\text{i}knt].$ 

Lastly $\kappa_{mk}={\rm Im}[k_{2 m}]$ are the imaginary parts of degree two, azimuthal order $m$ tidal Love numbers characterizing the primary's response to tidal driving at the (rotating frame) frequency $\omega_{mk}=kn-m\Omega$. In quasi-linear theory, tidal Love numbers can be computed through direct solution of the (Fourier-transformed) linearized fluid equations \citep[e.g.,][]{Ogilvie2013}. Alternatively, if the primary is a fluid planet or star with relatively simple structure, the response functions $\kappa_{mk}$ can be computed through an expansion in its normal mode oscillations \citep{Schenk2001,Braviner2015}. 

\subsection{Torque balance}
In any situation where the orbital angular momentum is significantly larger than the primary's spin angular momentum, \autoref{eq:Jdot} leads to a rapid adjustment of $J$ until the torque $\mathcal T$ vanishes. \citet[][hereafter D24]{Dewberry2024} explored spin evolution in a simple model of a planet or star with a uniform (but time varying) rotation rate $\Omega$. \citetalias{Dewberry2024} emphasized that if such an object possesses normal mode oscillations with frequencies $\omega$ that scale directly with $\Omega$ in the rotating frame, so that $\omega\approx\beta\Omega$ for some constant $\beta$, then vanishing torques at tidal frequencies $\omega_{mk}=kn-m\Omega\approx\omega$ can produce attractor states at particular ratios between $\Omega$ and the orbital mean motion:
\begin{equation}\label{eq:Omratio}
    \frac{\Omega}{n}
    \approx\frac{j}{m+\beta},
\end{equation}
where $j$ is one of the integers $k$ appearing in the summations of Equations \eqref{eq:P}-\eqref{eq:T}. 

This commensurability results from a balance between the torque from a non-wavelike, nearly hydrostatic ``equilibrium tide,'' and an opposing torque due to the near-resonantly driven mode. The scaling $\beta=\omega/\Omega$ of mode frequency with rotation rate is important because it leads to automatically adjusting spin evolution, ``trapping'' the primary's rotation rate into such a torque balance. The global inertial modes of fully convective planets or stars \citep{Braviner2014}, along with the inertial mode-like flows of partially convective bodies \citep{Lin2021} generally satisfy this frequency scaling criterion. \citetalias{Dewberry2024} used numerical calculations for a sequence of rotating, isentropic polytropes to demonstrate that inertial modes can arrest the evolution of stellar rotation rates toward the pseudosynchronization predicted weak frictional theory \citep{Hut1981}. 

\subsection{Eccentricity and semimajor axis evolution}\label{sec:eavol}
\citetalias{Dewberry2024} ignored orbital evolution, holding orbital eccentricity and semimajor axis fixed. To understand how they might evolve during an inertial mode torque balance, consider the decomposition 
\begin{equation}
    \mathcal{T}
    \simeq\mathcal{T}_j 
    +\mathcal{T}_{\rm hs}
    +\mathcal{T}_{\rm ext}
    \approx0,
\end{equation}
where $\mathcal{T}_j$ is the torque due to near resonant driving of an $m=2$ inertial mode by the $j$'th tidal harmonic, $\mathcal{T}_{\rm hs}$ is the torque associated with convective damping of a roughly hydrostatic tidal bulge, and $\mathcal{T}_{\rm ext}$ comprises any additional torques applied to the primary star (due to, e.g., magnetic braking; see Section \ref{sec:mbrake}). Ignoring $\mathcal{T}_{\rm ext}$ for the moment, note that the near-resonantly driven inertial mode's contribution to the tidal power is $\mathcal{P}_j=(jn/2)\mathcal{T}_j.$ Torque balance therefore leads to 
\begin{equation}
    \mathcal{P}
    \simeq
    \mathcal{P}_{\rm hs}
    +\mathcal{P}_i
    \simeq
    \mathcal{P}_{\rm hs} 
    -(jn/2)\mathcal{T}_{\rm hs},
\end{equation}
where $\mathcal{P}_{\rm hs}$ is the non-wavelike tidal power. Analytical expressions for $\mathcal{P}_{\rm hs}$ and $\mathcal{T}_{\rm hs}$ can be found under the additional approximation that all of the response coefficients $\kappa_{mk}^{\rm hs}$ associated with the non-wavelike tide have a linear dependence on tidal frequency. Writing $\kappa_{mk}^{\rm hs}=\tau\omega_{mk}$, where---for the sole purpose of illustration---$\tau$ is assumed to be a constant time lag shared by all tidal components, \autoref{eq:edot} for eccentricity evolution can be approximated as 
\begin{widetext}
    \begin{align}\label{eq:tbedot}  
        \frac{\dot{e}}{e}
        &\simeq
        -\frac{(1-e^2)}{4e^2}
        \epsilon_Tn^2\tau
        \left(
            H_{20} + 3H_{22}
            -3jH_{12} 
            -\frac{6jH_{12}}{2+\beta}
            +\frac{6j^2H_{02}}{2 + \beta}
        \right),
    \\\notag 
        &=-\frac{3}{2}
        \epsilon_Tn^2\tau
        \frac{(1-e^2)}{e^2}
        \left\{
            2 
            +j\frac{(j - 4 - \beta)}{2 + \beta}
            +\left[ 
                46
                +j\frac{(15j - 108 - 27\beta)}{2(2 + \beta)}
            \right]e^2
             +\mathcal{O}(e^4)
        \right\},
    \end{align}
\end{widetext}
where the functions $H_{pm}(e)=\sum_kk^p|X_{mk}|^2$ denote series expansions in Hansen coefficients that converge to polynomials in eccentricity (cataloged for convenience in \autoref{app:hansum}). 

\begin{figure*}
    \centering
    \includegraphics[width=0.67\linewidth]{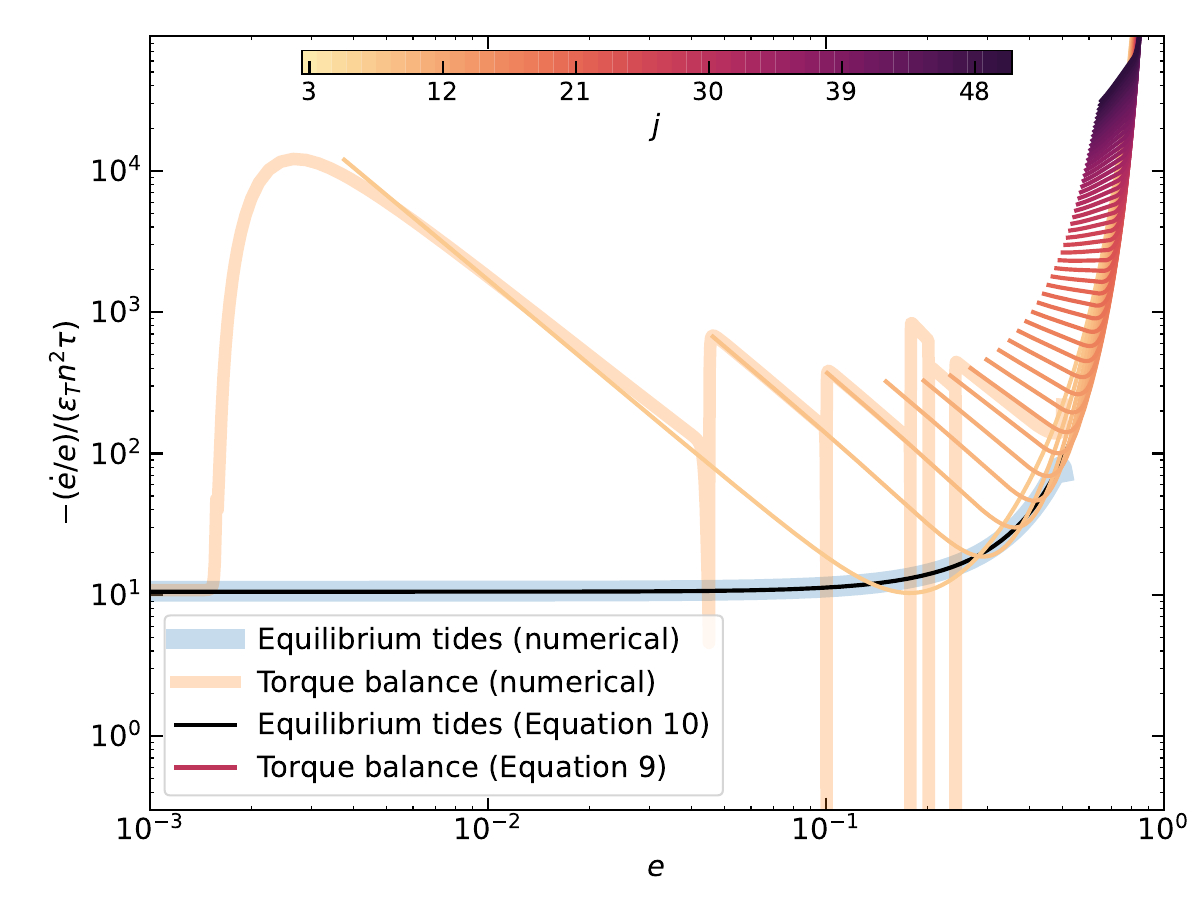}
    \caption{Inverse eccentricity damping timescales $\dot{e}/e,$ normalized by dimensionless amplitude $\epsilon_T=q(R/a)^5$, squared mean motion $n,$ and effective constant time lag $\tau$. The thin lines show the predictions of Equations \eqref{eq:psedot} and \eqref{eq:tbedot} for evolution due to equilibrium tides (black), and torque balances involving inertial modes (purple to orange colorscale). For the latter, each line is truncated at the point where Equation \eqref{eq:tbreak} predicts that the torque balance should break. The thick curves show results from directly integrating Equations \eqref{eq:adot}-\eqref{eq:Jdot} for a polytropic model of a $1M_\odot$ star with a reference radius $R_0=2R_\odot$, a constant kinematic viscosity of $\nu_0=10^{-5}R_0^2\omega_{\rm dyn,0}$, and the initial parameters $P_{\rm orb}=10$ days, $\Omega/\omega_{\rm dyn}=0.1,$ $e=0.5$ at $t=0$ (solutions computed from other initial conditions follow very similar trajectories in this parameter space). Tidally driven inertial modes can enhance eccentricity damping by orders of magnitude, relative to equilibrium tides.
    }
    \label{fig:edot_cpr}
\end{figure*}

\autoref{eq:tbedot} alone is not particularly useful, absent knowledge of appropriate values for the parameterized time lag $\tau$. However, it provides a direct point of comparison for the predictions of the constant time lag model of \cite{Hut1981}. Under the assumption that torque balance occurs at the ``pseudosynchronous'' ratio $\Omega/n=H_{12}/(2H_{02})$, purely constant time lag theories predict that tides damp eccentricity according to \citep[e.g.,][]{Leconte2010}
\begin{align}\label{eq:psedot}
    \frac{\dot{e}}{e}
    &\simeq
    -\frac{(1 - e^2)}{4e^2}\epsilon_Tn^2\tau
    \left(
        H_{20} + 3H_{22}
        -3\frac{(H_{12})^2}{H_{02}}
    \right)
    ,
\\\notag
    &=-\frac{21}{2}\epsilon_Tn^2\tau(1 - e^2) + \mathcal{O}(e^4).
\end{align}
\autoref{fig:edot_cpr} compares the righthand sides of \autoref{eq:tbedot} (thin colored curves) and \autoref{eq:psedot} (black curve), normalized by the common factor of $\epsilon_Tn^2\tau$, and adopting the frequency scaling $\beta=\omega/\Omega\simeq 0.7$ \citep[this value of the mode frequency ratio is typical for the longest wavelength, prograde inertial mode of both polytropic and more realistic stellar models; e.g.,][]{Dewberry2022}. The faint, thick colored curves plot the result of directly integrating Equations \eqref{eq:adot}-\eqref{eq:Jdot} using an $n=1.5$ polytropic model with parameters typical for a pre-main-sequence, solar-type star.

\autoref{fig:edot_cpr} demonstrates first that the tidal eccentricity damping found by directly integrating Equations \eqref{eq:adot}-\eqref{eq:Jdot} closely follows the predictions of Equations \eqref{eq:tbedot} and \eqref{eq:psedot}. Secondly, \autoref{fig:edot_cpr} shows that torque balance with inertial modes can drive eccentricity damping on timescales that are orders of magnitude faster than classical weak frictional theories. 

Physically, this boost in efficiency comes from the inertial modes' enhancement to tidal energy dissipation, which remains nonzero despite a nearly vanishing tidal torque. Mathematically, the order $e^2$ expression given in the second line of Equation \eqref{eq:psedot} demonstrates that in a constant time lag model $\dot{e}/e$ asymptotes to a constant as $e\rightarrow0$. In constrast, during a torque balance $\dot{e}/e$ scales like $e^{-2}$ as $e$ decreases. 

\subsection{Breaking torque balance}\label{sec:break}
The scaling $\dot{e}/e\propto e^{-2}$ must break down eventually, as the eccentric orbit transitions to a circular one. Ignoring external torques, this breakdown occurs when the amplitude of the $j$'th tidal harmonic is too small for the inertial mode torque $\mathcal{T}_j$ to counter-balance $\mathcal{T}_{\rm hs}.$ Quantitatively, torque balances break at the eccentricity value where
\begin{align}\label{eq:tbreak}
    \frac{2\pi|Q_2^\beta|^2}{5\epsilon_\beta\gamma_\beta}
    |X_{2k}|^2
    &=\frac{-\mathcal{T}_{\rm hs}}{3E_o\epsilon_T}
    \simeq 
    \tau n\left(
        H_{12} - \frac{2jH_{02}}{2 + \beta}
    \right),
\end{align}
where $Q_2^\beta$, $\epsilon_\beta,$ and $\gamma_\beta$ are constant properties of the balancing inertial mode (its overlap integral, energy constant, and damping rate, respectively; see \autoref{app:modes}). The left and righthand sides of this equation depend strongly on eccentricity (through $X_{2k},$ $H_{12}$, and $H_{02}$; see \autoref{app:hansum}), and weakly on semimajor axis (through the mean motion $n$). In \autoref{fig:edot_cpr}, the colored curves for each tidal harmonic $j$ terminate at the eccentricity where \autoref{eq:tbreak} is satisfied, assuming a fixed semimajor axis. The direct integration of Equations \eqref{eq:adot}-\eqref{eq:Jdot} plotted with the thick faint orange line demonstrates that these breaking points accurately predict where one torque balance breaks and transitions to another. 

\autoref{fig:edot_cpr} and Equations \eqref{eq:tbedot}-\eqref{eq:psedot} clearly indicate that the influence of tidally driven inertial modes on spin evolution may in turn play a critical role in damping orbital eccentricity. The rest of this paper presents simplified but physical calculations aimed at quantitatively estimating the importance of this eccentricity damping. 

\section{Numerical models and methods}\label{sec:model}
This section summarizes the essential details of the models we adopt in our numerical investigations of eccentricity damping by tidally driven inertial waves. 

\subsection{Internal structure models}\label{sec:eqm}
The interior model with the simplest wave response considered in this paper is a sequence of rotating polytropes with $P\propto\rho^{5/3}.$ Together with a first adiabatic exponent $\Gamma_1=5/3$, these polytropes provide a strictly isentropic approximation to planets or stars that are fully convective or have very small radiative or solid cores. The advantage of isentropic polytropes is that the subset of their inertial oscillations that respond appreciably to tidal driving is relatively sparse \citep[e.g.,][]{Braviner2014,Xu2017}. Appendix B in \citetalias{Dewberry2024} presents this sequence of rotating, centrifugally flattened models. 

We also consider tides raised in a simplified, homogeneous and incompressible fluid envelope lying on top of a solid, impermeable core. Such models roughly capture the behavior of tidally driven inertial waves when stable stratification in a star's radiative core is very strong \citep[e.g.,][]{Ogilvie2004,Ogilvie2007}. An incompressible fluid shell with a solid core has also historically been taken as a reasonable model for giant planets \citep{Ogilvie2009}. The advent of dilute and fuzzy cores models of Jupiter and Saturn \citep{Mankovich2021,Militzer2022} make this application less appropriate, since the buoyancy frequencies inferred for the interiors of these planets are low enough to allow mixing between inertial and gravito-inertial waves \citep{Lin2023,Dewberry2023,Pontin2024,Dhouib2024}. Solid core models may nevertheless capture the essential features of the tidal response of more realistic planets, if the buoyancy frequency is large relative to tidal frequencies \citep{Pontin2024}.

\begin{figure*}
    \centering
    \includegraphics[width=\linewidth]{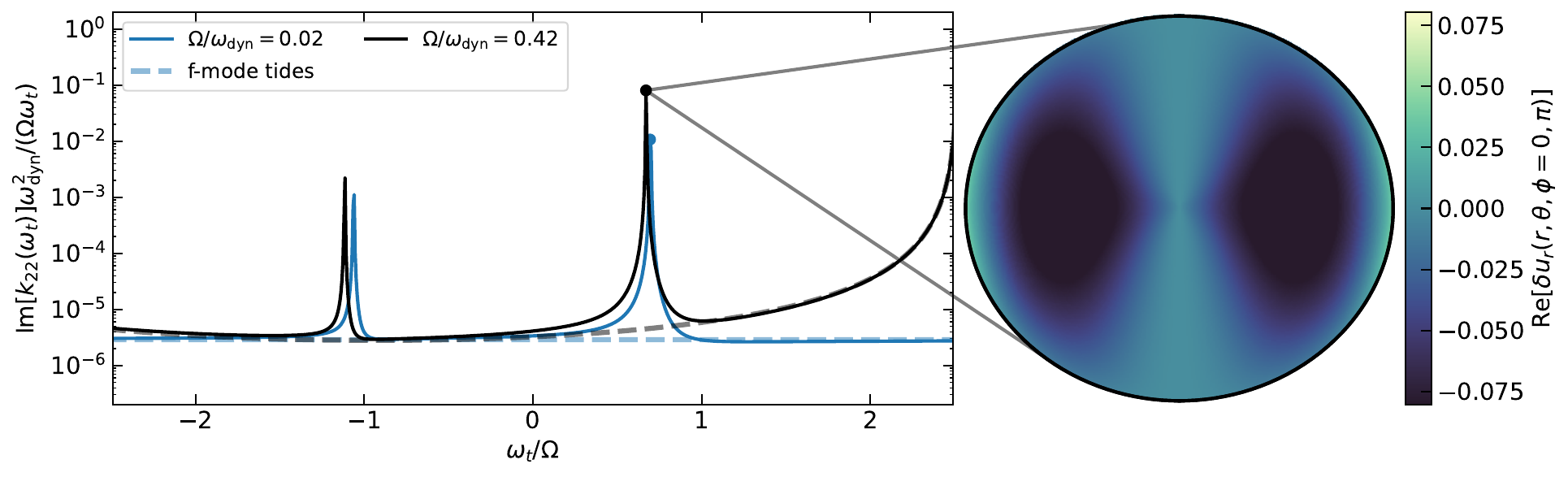}
    \includegraphics[width=\linewidth]{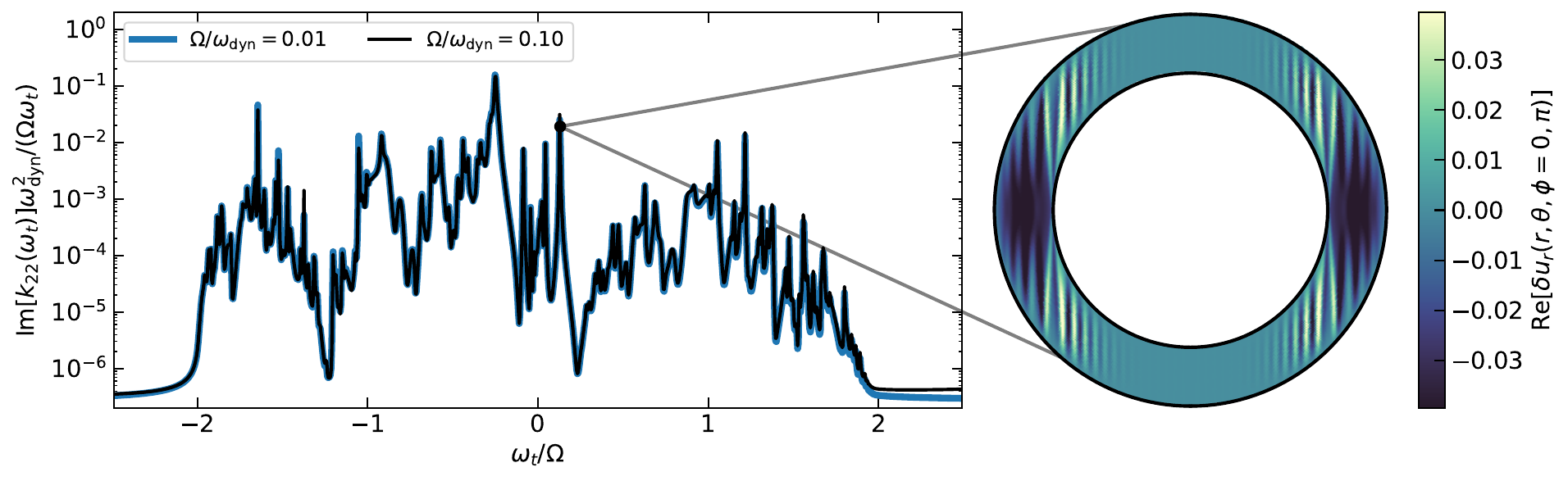}
    \caption{Left: imaginary parts of tidal Love numbers, scaled by $(\Omega/\omega_{\rm dyn})^2\omega_t/\Omega$ and plotted as a function of tidal frequency for isentropic polytropes (top) and incompressible shells (bottom) with a constant Ekman number $E_k=\nu/(R^2\Omega)=10^{-6}.$ The different curves show different rotation rates for the tidally perturbed object. The dashed lines in the top panel indicate ``f-mode tide'' calculations that exclude the inertial oscillations of the polytropic model. Right: meridional slices illustrating the radial velocity perturbation (in units with $G=M=R=1$, and normalized by the amplitude of the tidal potential) corresponding to the tidal frequency indicated by the black dots in the lefthand panels. The presence of a central core produces a much more complicated spectrum of resonances in the incompressible shell.
    }
    \label{fig:demo_iks}
\end{figure*}

Ignoring centrifugal distortion, incompressible shell models are characterized entirely by choices of the fractional dimensionless core radius $\alpha=r_{\rm core}/R$, the ratio $\rho_E/\overline{\rho}$ of envelope density $\rho_E$ to mean density $\overline{\rho}=M/(4\pi R^3/3)$, and the scaled rotation rate $\Omega/\omega_{\rm dyn}$ \citep[e.g.,][]{Ogilvie2009}. These piecewise homogeneous interior models are simpler than polytropes, admitting analytical solutions for the basic state rather than requiring numerical calculations. However, the inclusion of an impermeable core produces a much more complicated tidal flow (see \autoref{fig:demo_iks}, bottom). 

Throughout this paper, we adopt $M$, $R_0$, and $\omega_{\rm dyn,0}^{-1}=[R_0^3/(GM)]^{1/2}$ as units of mass, length, and time (respectively), where $R_0$ is the radius of the primary in the non-rotating limit. For our polytropic models, we allow the equatorial radius to vary with centrifugal flattening due to rotation as described in \citetalias{Dewberry2024}, while centrifugal flattening is ignored in our incompressible calculations.

\subsection{Eddy viscosity}\label{sec:nu}
Our calculations of tidal dissipation rates incorporate a viscosity to roughly describe damping by convective turbulence. We assume the functional form
\begin{equation}\label{eq:nudep}
    \nu=\nu_0(1 + C|\omega_t|^2/\omega_c^2)^{-1},
\end{equation}
where $\nu_0$ and $\omega_c$ are kinematic viscosities and convective turnover frequencies informed by mixing length theory, and $\omega_t$ is a rotating frame tidal frequency that in practice takes on values of $\omega_t=\omega_{mk}$. We consider calculations both with $C=0$ and with $C=0.876,$ the latter value providing a reasonable approximation to the frequency dependence inferred by \citet{Duguid2020}. We adopt values of $\nu_0/(R_0^2\omega_{\rm dyn,0})\simeq1-2\times10^{-5}$ and $\omega_c/\omega_{\rm dyn,0}\simeq10^{-4}-10^{-3}$ (specific values stated in connection with each set of calculations) that are motivated by mixing length theories for solar-like stars \citep[e.g.,][]{Ogilvie2007,Astoul2022,Dewberry2025}.

The non-monotonic dependence on frequency given by \autoref{eq:nudep} can play a similar role to inertial wave resonances, introducing additional spin equilibria where the tidal torque vanishes. This was previously recognized by \citet{Ivanov2004}, and has more recently played a role in analyses of tidal torques in viscoelastic planets with Maxwell rheologies \citep{Makarov2013,Storch2014,Valente2022}. Depending on the values assumed for $\nu_0$ and $\omega_c$, these torque balances can occur at ratios $\Omega/n\simeq k/m$ where the tidal frequency $\omega=kn - m\Omega$ of the $k,m$-component of the tidal potential itself vanishes (in turn producing a maximum in \autoref{eq:nudep}).

\subsection{Response function calculations}
The task of accurately computing the response functions $\kappa_{mk}$ that appear in Equations \eqref{eq:P}-\eqref{eq:T} for realistic models of planets and stars provides a significant hurdle to developing practical tidal theories, even when the amplitude of the tidally driven flow remains small enough for a quasilinear theory to be reasonable \citep{Ogilvie2004,Ogilvie2009,Ogilvie2013,Goodman2009,Ivanov2010}. Because of this, many studies of tides appeal to a parameterized constant ``quality factor'' or time lag, for which the functions $\kappa_{mk}$ are constant or vary linearly with tidal frequency. 

The calculations in this paper use linear response functions computed self-consistently from the background models described in Section \ref{sec:eqm}, and the eddy viscosity described in Section \ref{sec:nu}. The lefthand panels in \autoref{fig:demo_iks} plot example profiles of ${\rm Im}[k_{22}]$ as a function of tidal frequency $\omega_t$ for different models and rotation rates. The meridional slices shown on the right side of \autoref{fig:demo_iks} illustrate  the $(r,\theta)$ structure of the radial velocity perturbation driven at the frequencies indicated by the black dots. 

\subsubsection{Polytropic model}
For our polytropic models of fully convective stars (top panels of \autoref{fig:demo_iks}), we use the same mode expansion described in \citetalias{Dewberry2024}. This expansion comprises, for each azimuthal order $m$ in a quadrupolar tidal potential, the degree $\ell\simeq2$ fundamental oscillations, as well as the two (prograde and retrograde) inertial oscillations with the longest wavelengths. Appendix C in \citetalias{Dewberry2024} presents the properties of the $m=2$ subset of these oscillations, which are required to compute the coefficients $\kappa_{2k}.$ \autoref{app:modes} presents the properties of $m=0$ oscillations additionally needed to compute the coefficients $\kappa_{0k}$ appearing in \autoref{eq:P} for the tidal power $\mathcal{P}$. In both cases, the modes are computed from the inviscid linearized equations, adopting free-surface boundary conditions.

Limiting our mode expansion to such a small number of oscillations excludes resonances with a dense spectrum of shorter wavelength inertial modes \citep[e.g.,][]{Wu2005}. We expect the importance of these excluded modes to increase more and more in the limit of weaker and weaker damping. However, our results with the incompressible model introduced in the next section suggest that the presence of additional inertial oscillations should serve only to enhance the spin evolution and eccentricity damping produced here by their longest wavelength representatives.

A convenient feature of our polytropic model is that simply excluding inertial oscillations from the mode expansion produces a very close analogue to equilibrium tides, at least in the low-frequency regime \citep{Dewberry2025}. We refer to this approximation (plotted with the dashed lines in the top panel of \autoref{fig:demo_iks}) as ``f-mode tides,'' and use it as a point of comparison. 

\subsubsection{Incompressible model}
For our incompressible shell models of convective envelopes (bottom panels of \autoref{fig:demo_iks}), we directly compute the response functions $\kappa_{mk}$ using the pseudospectral method described in \citet{Ogilvie2009}. These calculations adopt stress-free, free-surface boundary conditions at the surface, and stress-free, impenetrable conditions at the central core.

Directly solving the Fourier-transformed problem is more numerically expensive than approximating the tidal response with a mode expansion. For our incompressible model, we therefore pre-compute ${\rm Im}[k_{2m}]\omega_{\rm dyn}^2/(\Omega\omega_t)$ values\footnote{
In incompressible models with a constant Ekman number and a free-surface boundary condition, this dimensionless combination remains an approximately invariant function of $\omega_t/\Omega$ under changes in $\Omega$, as long as $\omega_t\ll\omega_{\rm dyn}$ (compare the blue and black curves in the bottom left panel of \autoref{fig:demo_iks}).
} on a large grid of tidal frequencies (with $1200$ evenly spaced points between $\pm2\Omega$), from which we interpolate the required coefficients $\kappa_{2m}\omega_{\rm dyn}^2/(\Omega\omega_{mk})$ during integration of the secular equations.

\subsection{Magnetic braking}\label{sec:mbrake}
To estimate the effect of magnetic braking on inertial mode torque balance, we include a braking torque with the form \citep{Skumanich1972,Verbunt1981,Rappaport1983,Paxton2015}
\begin{equation}\label{eq:mbrake}
    T_B
    =-1.6\times10^{36}
    \hat{I}
    \left(\frac{M}{M_\odot}\right)
    \left(\frac{R}{R_\odot}\right)^4
    \left(\frac{1 {\rm \ d}}{P_{\rm rot}}\right)^3
    [{\rm dyn \ cm}],
\end{equation}
where $P_{\rm rot}=2\pi/\Omega$ is the primary body's rotation period, $\hat{I}=I/(MR^2)$ is its dimensionless moment of inertia, and we assume $M_\odot=1.9884\times10^{33}{\rm g},$ 
$R_\odot=6.957\times10^{10}{\rm cm}.$

\subsection{Numerical integrations}
Both forms of interior and tidal model yield response functions $\kappa_{mk}$ that can be used to compute the tidal power and torque, and in turn incorporated into the time integration of Equations \eqref{eq:adot}-\eqref{eq:Jdot}. We primarily integrate these equations using \texttt{scipy}'s \texttt{solve\_ivp} (with \texttt{method='LSODA',rtol=1e-4}). Abrupt variations in the torque make these equations somewhat stiff, so we have checked that the Gragg-Bulirsch-Stoer integrator included in \texttt{REBOUND} \citep{rebound} produces the same results. 

\begin{figure}
    \centering
    \includegraphics[width=\columnwidth]{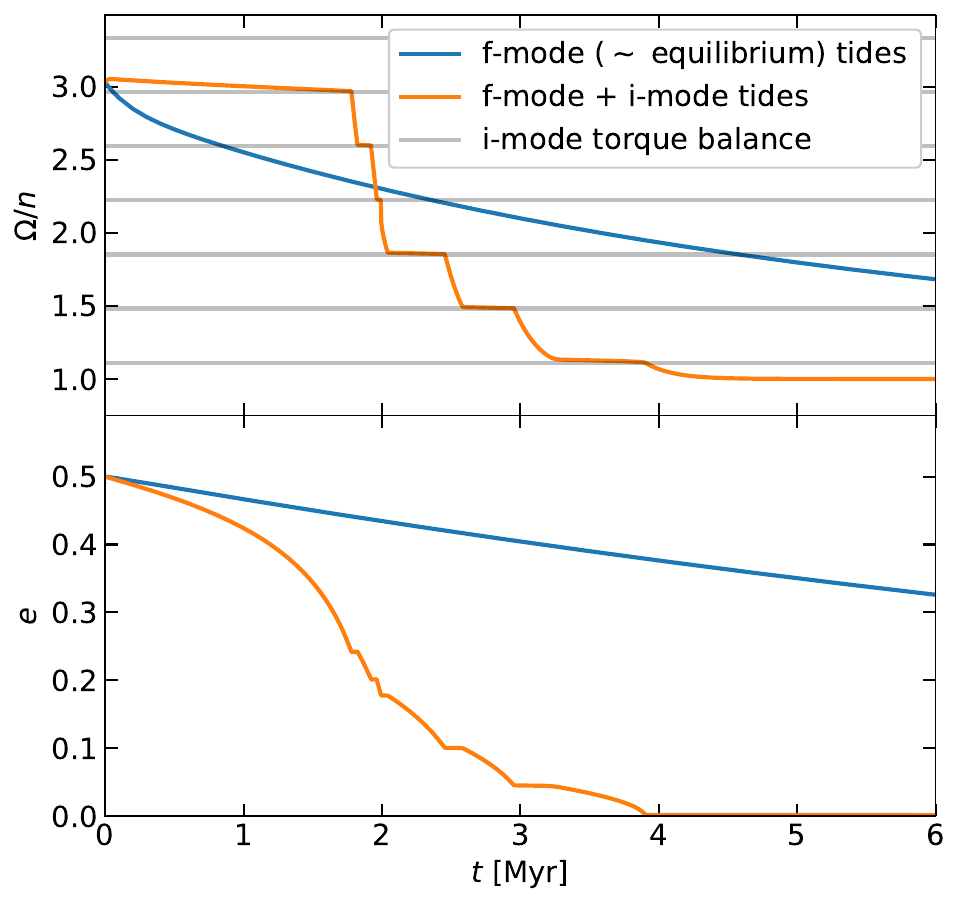}
    \caption{Time evolution of the ratio between rotation rate and mean motion (top) and eccentricity (bottom) for the integrations depicted by thick lines in \autoref{fig:edot_cpr}. In calculations including the inertial modes of the model (orange lines), a progression through discrete spin ratios $\Omega/n\simeq j/2.7$ (for integer $j$; gray lines) leads to much more rapid eccentricity damping than in an equilibrium tidal model with the same underlying eddy viscosity (taken here to have a constant value of $\nu=10^{-5}R_0^2\omega_{\rm dyn,0},$).
    }
    \label{fig:demo_ivf}
\end{figure}

\begin{figure}
    \centering
    \includegraphics[width=\linewidth]{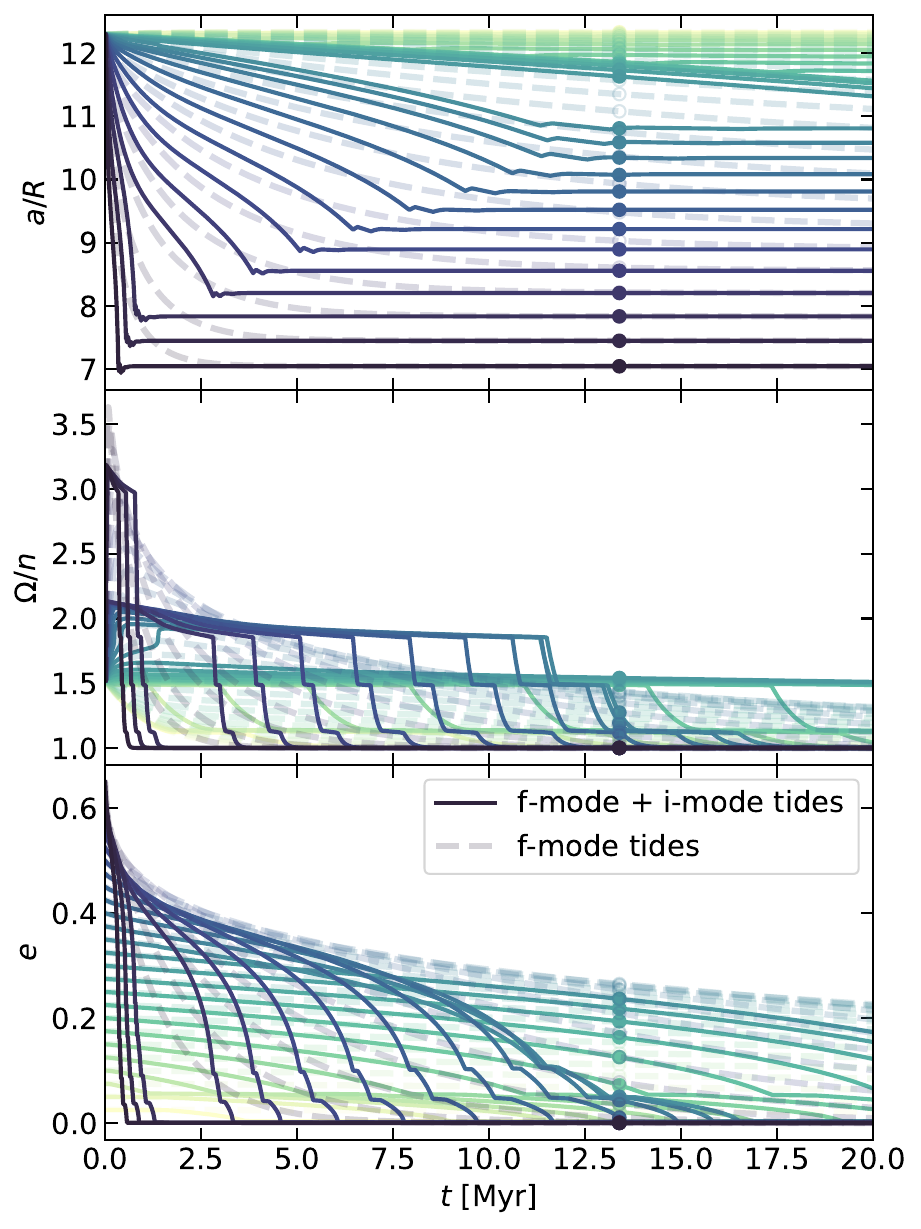}
    \caption{Evolution of semimajor axis divided by equatorial radius (top), period ratio $\Omega/n=P_{\rm orb}/P_{\rm rot}$ (middle), and eccentricity (bottom), for polytropic models with $M=1M_\odot$, $R=2R_\odot$, constant kinematic viscosity $\nu=10^{-5}R_0^2\omega_{\rm dyn,0},$ mass ratio $q=1,$ initial orbital periods and rotation rates of $P_{\rm orb}=10$ days and $\Omega/\omega_{\rm dyn}=0.05$ (respectively), and initial eccentricities increasing from 0 (light yellow) to $0.65$ (dark blue). The dots at $t\simeq13$Myr indicate the simultaneous endpoints for the parameterized curves shown in \autoref{fig:n15_eaO} (left). Larger initial eccentricities lead to more rapid circularization, regardless of the tidal model. For any given initial condition, tidally driven inertial modes (solid lines) damp eccentricities much more rapidly than quasi-equilibrium tides (dashed lines).
    }\label{fig:n15_aoe_vs_t}
\end{figure}

\section{Results}\label{sec:res}

\subsection{Inertial waves vs. weak frictional tides}
\autoref{fig:demo_ivf} demonstrates the impact of tidal inertial mode driving in the simplest approximation. The panels plot the evolution of the spin (top) and eccentricity (bottom) for a polytropic model of a solar mass star on the pre-main-sequence with an equal mass companion. Starting from exactly the same initial conditions and adopting an otherwise identical interior model, calculations including inertial modes (orange) contrast sharply with those that omit them (blue). Tidal driving of the longest wavelength prograde inertial mode produces a cascade through spin equilibria with $\Omega/n\simeq j/2.7$ (gray lines), where $j$ is the integer index of the tidal harmonic responsible for driving the mode at a given point in time. The spin remains locked in the $j$'th ratio with the mean motion until the eccentricity damps to the point where the torque balance breaks (see Section \ref{sec:break}). Comparing the top and bottom panels of \autoref{fig:demo_ivf}, the eccentricity damping takes place almost exclusively during these locked states (in fact, the eccentricity commonly increases marginally during spin transitions).

\begin{figure*}
    \centering
    \includegraphics[width=\linewidth]{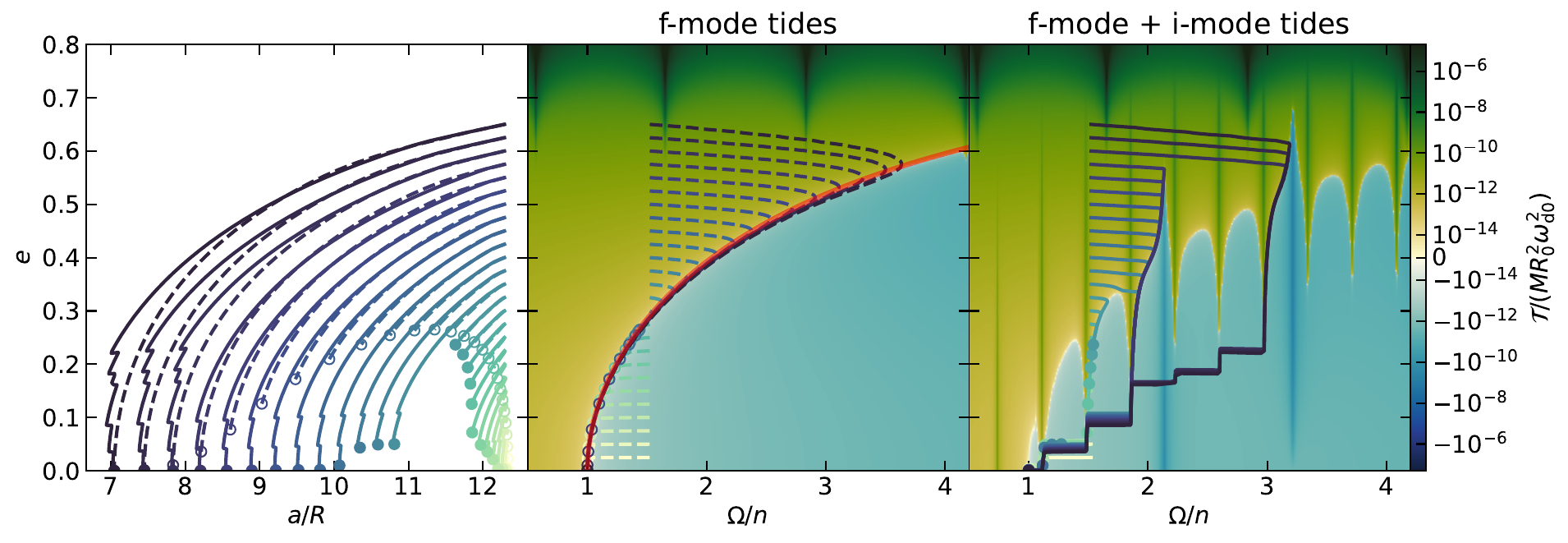}
    \caption{Parameterized curves showing the same simulations as \autoref{fig:n15_aoe_vs_t}, but with eccentricity plotted as a function of $a/R$ (left) and $\Omega/n$ (middle, right). The middle and righthand panels demonstrate that the evolution progresses along the surface of vanishing tidal torque (indicated by the colormap, which transitions from log to linear scale near $\mathcal{T}=0$). The red line in the middle panel shows the zero-torque surface predicted for a constant time lag. The torque is qualitatively modified by the inclusion (right) or omission (middle) of inertial waves in the tidal model. The lefthand panel shows that inertial modes' modification of tidal energy dissipation produces circularization out to larger $a/R$ in an equivalent amount of time ($t\approx13$Myr).
    }
    \label{fig:n15_eaO}
\end{figure*}

The overall impact of the tidally driven inertial modes is to significantly shorten the circularization time. The bottom panel of \autoref{fig:demo_ivf} demonstrates that torque balancing inertial modes can damp an initially moderate eccentricity to nearly zero much faster than equilibrium tidal models adopting the same input eddy viscosity. \autoref{fig:n15_aoe_vs_t} extends this comparison to a wider range of initial eccentricities (indicated by line color). For a given initial condition, eccentricity damping by torque balanced inertial modes (solid curves) far outstrips eccentricity damping by f-mode tides (dashed curves). 

\autoref{fig:n15_eaO} translates the time integrations from \autoref{fig:n15_aoe_vs_t} to a more physically meaningful (and observable) parameter space. The left panel in \autoref{fig:demo_ivf} plots the evolution of eccentricity against the ratio of semimajor axis to (equatorial) stellar radius. Over the same amount of time, inertial mode driving circularizes out to significantly larger values of $a/R$ than more classical models of equilibrium tides (the open/filled points in the left panel of \autoref{fig:n15_eaO} correspond to those shown in \autoref{fig:n15_aoe_vs_t}). 

\begin{figure*}
    \centering
    \includegraphics[width=\linewidth]{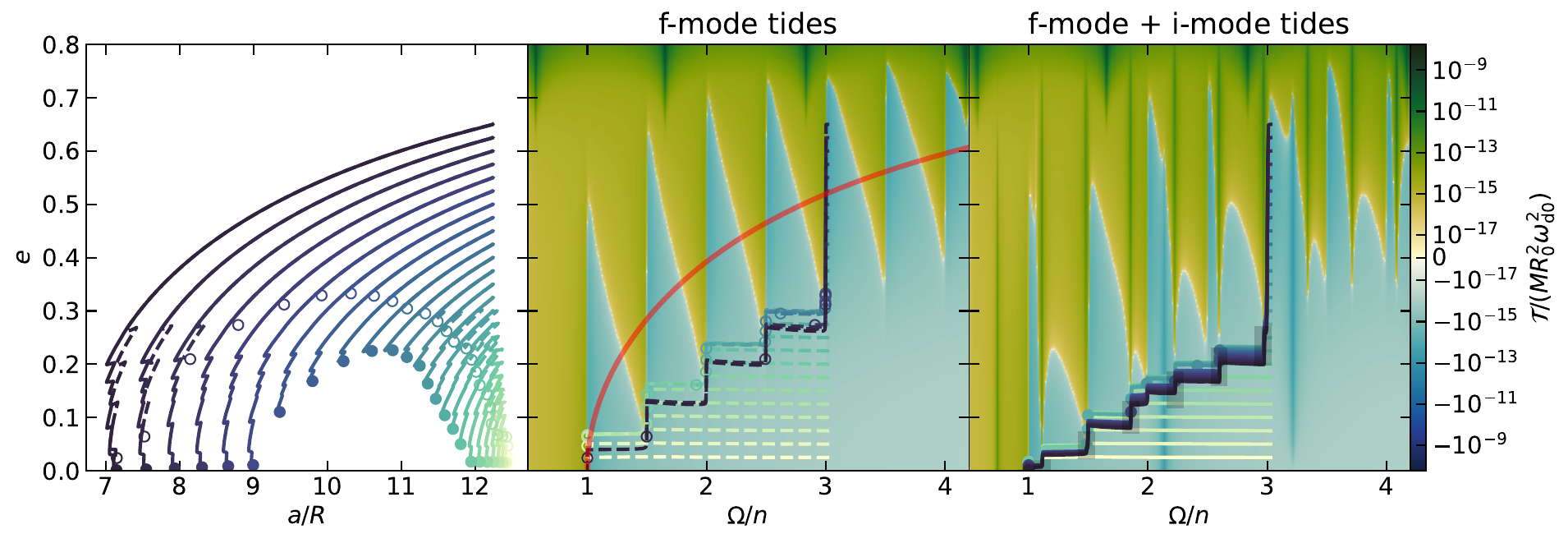}
    \caption{Simulations run with similar parameters to Figures \ref{fig:n15_aoe_vs_t} and \ref{fig:n15_eaO}, but starting from $\Omega/\omega_{\rm dyn}=0.10$ and including high-frequency suppression of the effective eddy viscosity (with values of $\nu_0=10^{-5}R_0^2\omega_{\rm dyn,0}$, $C=0.876$, and $\omega_c=10^{-3}$ incorporated into \autoref{eq:nudep}). High-frequency suppression introduces similar torque balances even without inertial modes (vertical ridges in the middle panel), at spin ratios $\Omega/n=k/2$ for integer $k=2,3,4,...$ (see Section \ref{sec:nu}). The difference between calculations with and without tidally driven inertial waves are consequently less pronounced, as illustrated by the leftmost panel. In both cases, high-frequency suppression of the eddy viscosity greatly slows evolution (the points in the lefthand panel indicate $t\approx 20 $ Gyr!).
    }\label{fig:n15_eaO_omc}
\end{figure*}

The center and right panels in \autoref{fig:n15_eaO} plot eccentricity as a function of $\Omega/n.$ The tidal torque shown by the colormap in these panels demonstrates that tidal evolution progresses along the surfaces in parameter space where $\mathcal{T}\approx0$. In the case without inertial modes, evolution due to f-mode tides closely tracks the pseudosynchronous spin ratio predicted by \citet{Hut1981} (red curve). The rightmost panel in \autoref{fig:n15_eaO} demonstrates that inertial mode torque balances enhance eccentricity damping by driving spin evolution that deviates from this pseudosynchronization. 

\begin{figure*}
    \centering
    \includegraphics[width=\linewidth]{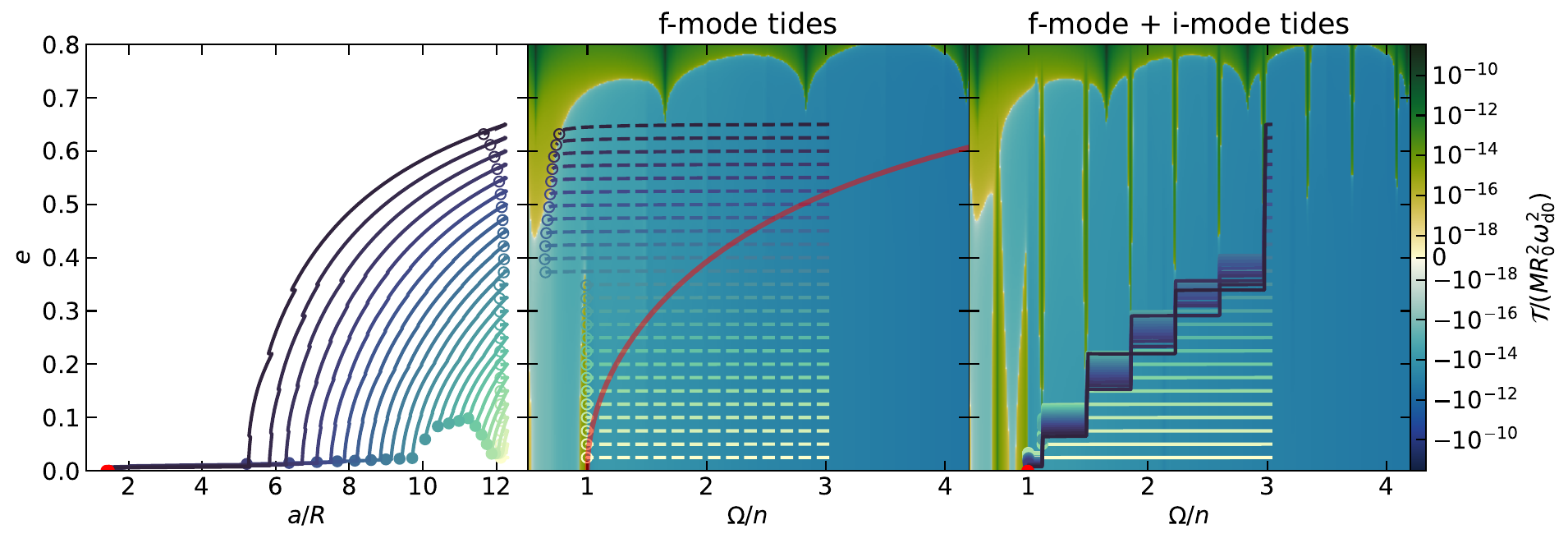}
    \caption{Simulations run with similar parameters to \autoref{fig:n15_eaO_omc}, but incorporating magnetic braking (via \autoref{eq:mbrake}). Red dots in the leftmost panel indicate simulations with a primary star expected to overfill its Roche lobe and begin mass transfer. The additional spin-down torque removes most of the spin equilibria produced solely by high-frequency suppression of the eddy viscosity (compare the middle panel here with that of \autoref{fig:n15_eaO_omc}), but leaves behind those associated with inertial mode resonances (narrow ridges in the rightmost panel). With both high-frequency suppression and magnetic braking, inertial modes drive much more significant eccentricity damping than equilibrium tides, in this case over a period of $t\approx 830$Myr.
    }\label{fig:n15_eaO_omc_mb}
\end{figure*}

\subsection{High-frequency suppression}
Suppression of the effective viscosity at large tidal frequencies can make the differences between eccentricity damping with and without dynamically driven inertial modes less pronounced. \autoref{fig:n15_eaO_omc} shows similar calculations to those depicted in \autoref{fig:n15_eaO}, but with nonzero factors $C$ and $\omega_c$ included in \autoref{eq:nudep}. The middle panel demonstrates that the frequency dependence of the eddy viscosity itself introduces ratios $\Omega/n\simeq k/2$ where the total torque vanishes, and the spin evolution stalls \citep{Ivanov2004}. The combination of these ``viscous'' torque balances with inertial modes then produces a more complicated zero-torque surface (rightmost panel). 

Evolution traces the zero-torque surface in both cases, with eccentricity damping again taking place exclusively during torque balance. Tidally driven inertial modes still accelerate eccentricity damping relative to f-mode tides with the same viscosity, as shown in the leftmost panel of \autoref{fig:n15_eaO_omc}. However, the overall reduction of that viscosity at large tidal frequencies strongly slows evolution relative to the un-suppressed case shown in \autoref{fig:n15_aoe_vs_t}. The points in the lefthand panel indicate that even after $t\approx20$Gyr, neither f-mode tides nor inertial waves produce circularization out to $a/R=10$.

\subsection{Magnetic braking}
The impact of magnetic braking on orbital and spin evolution depends on the amplitude of the magnetic braking torque (\autoref{eq:mbrake}) relative to that of the tidal torque. In the absence of high-frequency suppression, magnetic braking does very little to change the evolution described by \autoref{fig:n15_eaO}, since the braking torque is always much less than the tidal torque for the values of $M=M'=1M_\odot$, $R_0=2R_\odot,$ and $\nu_0=10^{-5}R_0^2\omega_{\rm dyn,0}$ adopted.

\begin{figure}
    \centering
    \includegraphics[width=\linewidth]{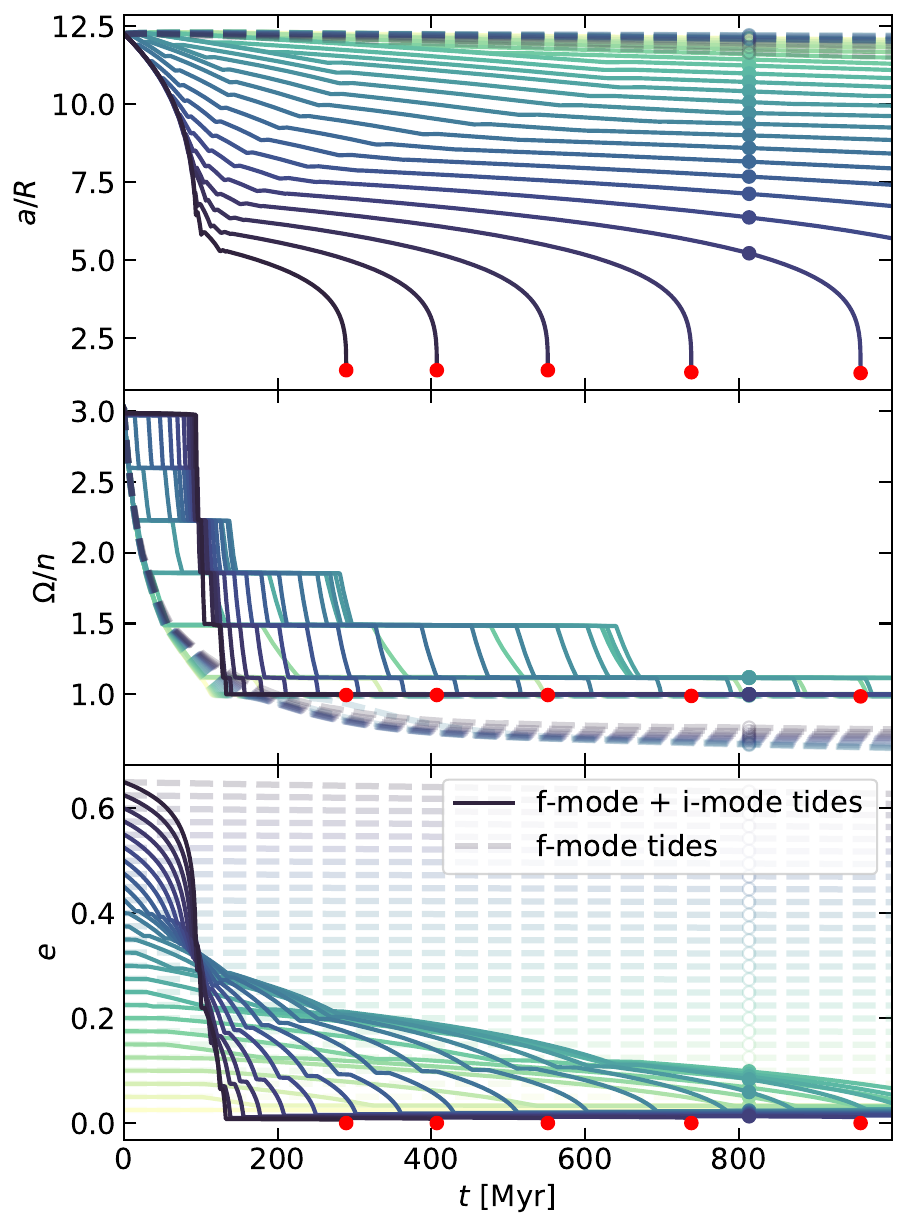}
    \caption{Time evolution for the simulations shown in \autoref{fig:n15_eaO_omc_mb}. Although inertial modes still produce much more rapid eccentricity damping than purely equilibrium tides, high-frequency suppression of the eddy viscosity significantly slows tidal evolution (compare with \autoref{fig:n15_aoe_vs_t}). 
    }\label{fig:n15_aoe_vs_t_mb}
\end{figure}

When the eddy viscosity is suppressed at large tidal frequencies, however, magnetic braking can have a much more significant effect. \autoref{fig:n15_eaO_omc_mb} and \autoref{fig:n15_aoe_vs_t_mb} summarize calculations including both high-frequency suppression and magnetic braking. The magnetic braking torque is large enough in comparison with the frequency-suppressed tidal torque to entirely erase the viscous torque balances from \autoref{fig:n15_eaO_omc} (middle). Inertial mode torque balances remain apparent in the right panel of \autoref{fig:n15_eaO_omc_mb} because the modes are weakly damped enough that their resonances can still provide a counterweight for the spin-down torque. The lefthand panel of \autoref{fig:n15_eaO_omc_mb} describes an even more dramatic enhancement of the tidal eccentricity damping by inertial modes, compared to damping by equilibrium or non-wavelike tides. However, the curves in \autoref{fig:n15_aoe_vs_t_mb} show that this orbital evolution still takes place over a much longer period of time than the un-suppressed, un-braked calculations shown in \autoref{fig:n15_aoe_vs_t}.

\subsection{Mode amplitudes and damping rates} 
\begin{figure}
    \centering
    \includegraphics[width=\columnwidth]{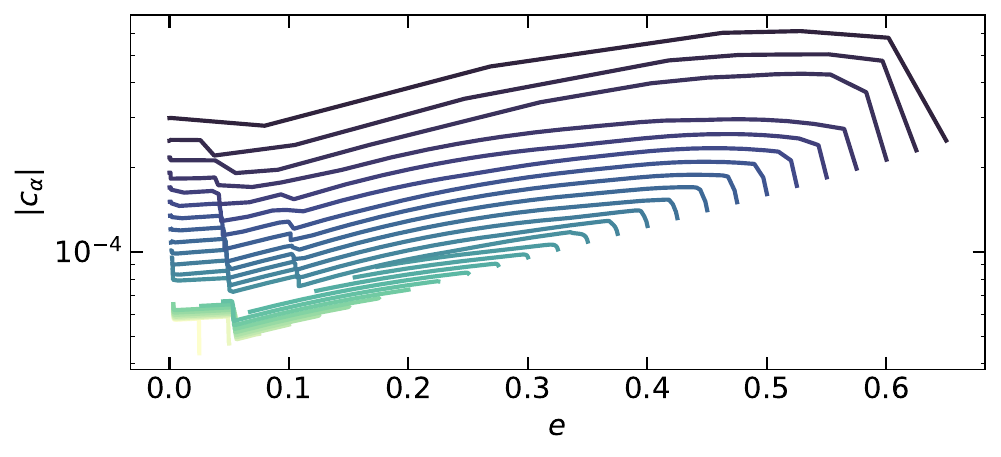}
    \includegraphics[width=\columnwidth]{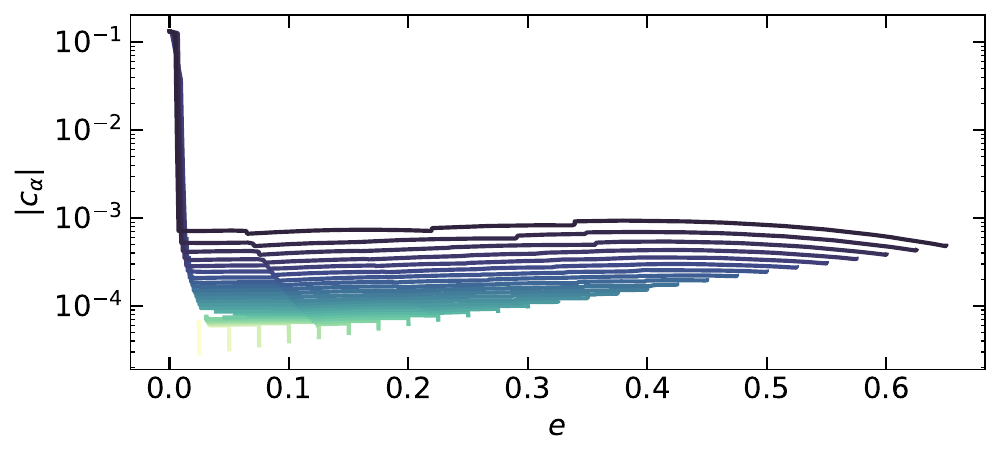}
    \caption{Amplitudes of the longest wavelength, prograde, $m=2$ inertial mode, computed (via \autoref{eq:camp}) for each of the simulations shown by solid lines in \autoref{fig:n15_eaO} (top panel) and \autoref{fig:n15_eaO_omc_mb} (bottom panel). Because the inertial modes are only driven strongly enough to counterbalance a relatively weak opposing torque, their amplitudes remain $\ll1$ unless decreasing separations drive the system toward mass transfer and coalescence (this is the case as $e\rightarrow0$ in the bottom panel).}
    \label{fig:n15p_amps}
\end{figure}

Tidal interactions involving sustained mode resonances commonly lead to complications as mode amplitudes become large. For example, internal gravity mode resonance locking combined with geometric focusing in the radiative cores of stars can induce nonlinear wave breaking \citep[e.g.,][]{Ma2021,Zanazzi2021}. Nonlinear wave breaking is not a major concern for the evolution shown in Figures \ref{fig:n15_aoe_vs_t}-\ref{fig:n15_eaO_omc_mb}, however. The top and bottom panels of \autoref{fig:n15p_amps} plot dimensionless amplitudes computed for the prograde $m=2$ inertial mode responsible for the evolution shown in lefthand panels of \autoref{fig:n15_eaO} and \autoref{fig:n15_eaO_omc_mb} (respectively). These amplitudes are computed as the absolute value of 

\begin{figure*}
    \centering
    \includegraphics[width=\textwidth]{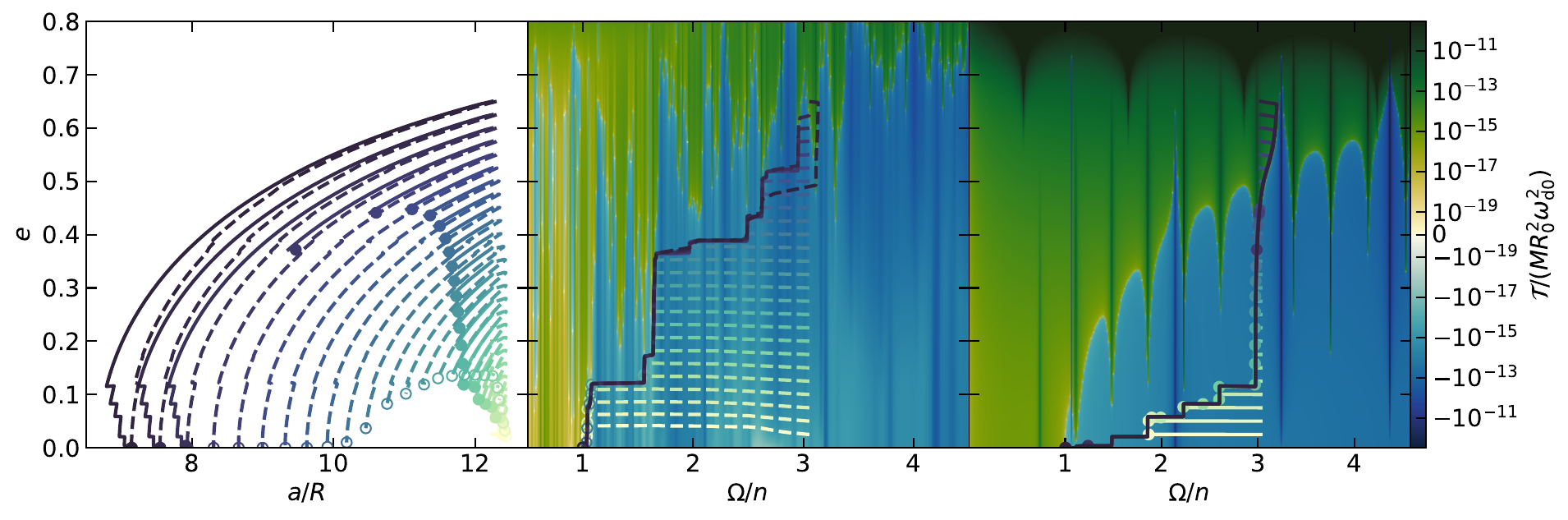}
    \caption{Comparison between tidal evolution for polytropic (solid lines, right panel) and incompressible shell (dashed lines, middle panel) models of a $1M_\odot,$ and initially $2R_\odot$ star. The fractional core radius for the incompressible shell is given by $r_C/R=0.7$, and the ratio of envelope to mean density is $\rho_E/\overline{\rho}=0.04.$ Each curve shows simulations with $q=1,$ and initial parameters $P_{\rm orb}=10$ days, $\Omega/\omega_{\rm dyn}=0.1,$ and $e\in[0,0.65]$. All of the simulations plotted assume a constant Ekman number of $E_k=\nu/(R^2\Omega)=10^{-6}$, and terminate at $t\approx83$Myr. With the addition of a central core, the inertial waves of the incompressible shell damp eccentricities even more rapidly than those of the fully convective polytrope, driving similar evolution but across a more complicated zero-torque surface.}
    \label{fig:inc_Ek6}
\end{figure*}
\begin{equation}\label{eq:camp}
    c_\beta
    \simeq
    \sum_{k=-\infty}^\infty \frac{4\pi q(R/a)^3 X_{2k} Q_2^\beta}
    {5\epsilon_\beta(\Delta_{\beta k2} - \text{i}\gamma_\beta)}
\end{equation}
where $\Delta_{\beta km}=\omega_\beta - (kn - m\Omega)$ is the detuning between the mode's rotating frame frequency and the frequency of the $k'$th tidal harmonic. 

The amplitudes shown in \autoref{fig:n15p_amps} are not large enough to make wave breaking a concern, at least for the long wavelength inertial mode in question. Magnetic braking increases the amplitudes marginally, since it provides an additional torque for the driven mode to counterbalance, but only after eccentricities have been damped away (as the star transitions to a mass transferring regime). 

Low mode amplitudes are consistent with the finding from \citetalias{Dewberry2024} that they should remain small and roughly independent of dissipative processes that act both on the torque balancing mode and the non-wavelike tide. Throughout this paper we have implicitly assumed that inertial waves and the non-wavelike tide are damped similarly by convection. Stronger damping (by e.g. Ohmic diffusion of magnetohydrodynamic inertial oscillations) could increase damping rates enough to preclude torque balance in the first place (i.e., by increasing the amplitude of $\gamma_j$ in the denominator of \autoref{eq:tbreak}). Alternatively, if inertial modes are much more weakly damped by convective eddies, small values of $\Delta_{\beta k2}$ and $\gamma_\beta$ in \autoref{eq:camp} could produce much larger mode amplitudes than those found here. Even without differences between damping of the wavelike and non-wavelike tide, nonlinear wave breaking is intrinsically more important for shorter wavelength inertial waves than the global inertial modes of isentropic polytropes considered so far in this paper.

\subsection{Influence of a central core}\label{sec:ccore}
\autoref{fig:inc_Ek6} compares results for a simplified polytropic star against evolution for an incompressible shell model of the primary's convective envelope. In the incompressible shell, a more complicated dependence of the tidal power and torque on tidal frequency (see \autoref{fig:demo_iks}, bottom) produces a more complicated zero-torque surface. Nevertheless, the overall scaling of inertial waves frequencies with $\Omega$ still leads to sustained torque balances and enhanced eccentricity damping. This enhancement in fact outstrips that exhibited by the simpler polytropic model: as shown in the lefthand panel of \autoref{fig:inc_Ek6}, over the same period of time ($t\approx 83 $Myr in this case) and for otherwise identical parameters, tides raised in the incompressible shell damp eccentricities out to a larger relative orbital separation.

We have not attempted to include high-frequency suppression in our incompressible shell calculations, since varying the Ekman number self-consistently would be much more numerically expensive \citep[qualitative changes to the spectrum of mode resonances appearing in the bottom panel of \autoref{fig:demo_iks} during evolution would preclude the interpolation used in this paper; e.g.,][]{Ogilvie2013}. The calculations shown in \autoref{fig:inc_Ek6} suggest that inertial waves in planets and stars with both convective envelopes and stably stratified or solid cores can drive similar eccentricity damping to the inertial modes of fully convective models. However, the efficiency of eccentricity damping exhibited by our fully convective polytropic model may need to be taken as a lower bound. 

\subsection{Orbital population synthesis}\label{sec:ps}
To explore how tidally driven inertial waves might affect observational data, we integrate Equations \eqref{eq:adot}-\eqref{eq:Jdot} for a population of $5000$ initial conditions motivated by observations of relatively short-period, solar-type binaries. We draw initial orbital periods from a uniform distribution between $P_{\rm orb}=1$ and $100$ days. We draw initial eccentricities and primary rotation periods both from Rayleigh distributions, with modes of $0.3$ and $15$ days, respectively. We then integrate the secular equations for $10^9$ years, using our polytropic model, with $M=1M_\odot$, $R_0=2R_\odot$, $\nu_0=2\times 10^{-5}R_0^2\omega_{\rm dyn,0}$, and $\omega_c$ fixed at either $5\times 10^{-4}\omega_{\rm dyn,0}$ or $5\times10^{-3}\omega_{\rm dyn,0}$. 

This model and set of parameters are motivated by pre-main-sequence evolution in the \texttt{1m\_pre\_ms\_to\_wd} suite of Modules for Experiments in Stellar Astrophysics \citep[MESA][]{Paxton2011, Paxton2013, Paxton2015, Paxton2018, Paxton2019, Jermyn2023}, although we omit stellar evolution from our simlutations (the equatorial radius evolves, but only with changing rotation). The values of $\nu_0$ and $\omega_c$ come from mass-weighted averages of $\ell_cv_c$ and $v_c/\ell_c$, where $\ell_c$ and $v_c$ are convective mixing lengths and velocities (respectively).

Using these parameters in models of isentropic polytropes run out to 1 Gyr is unrealistic, since the inertial wave response of a realistic solar-type star will change as it develops a radiative core. While the calculations presented in Section \ref{sec:ccore} suggest that the presence of a central core does not fundamentally alter the mechanism of torque balance (only its efficiency), the simulations presented in this section should only be expected to predict general trends.

\subsubsection{Circularization}
\begin{figure*}[h!]
    \centering
    \includegraphics[width=\textwidth]{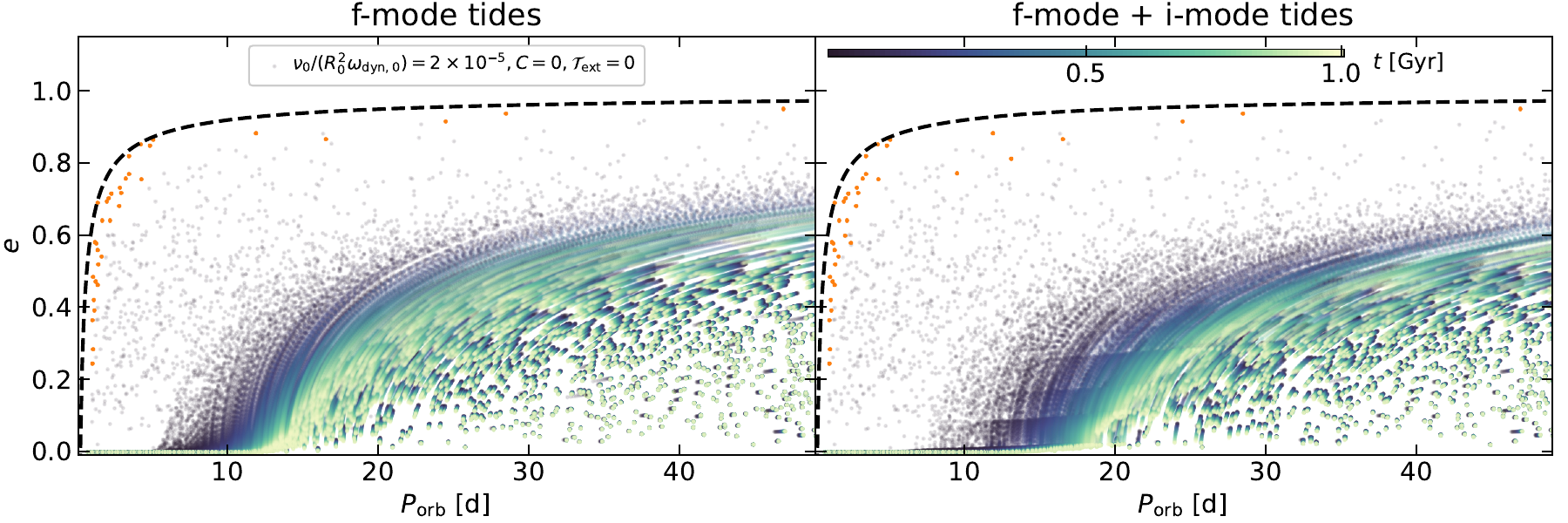}
    \includegraphics[width=\textwidth]{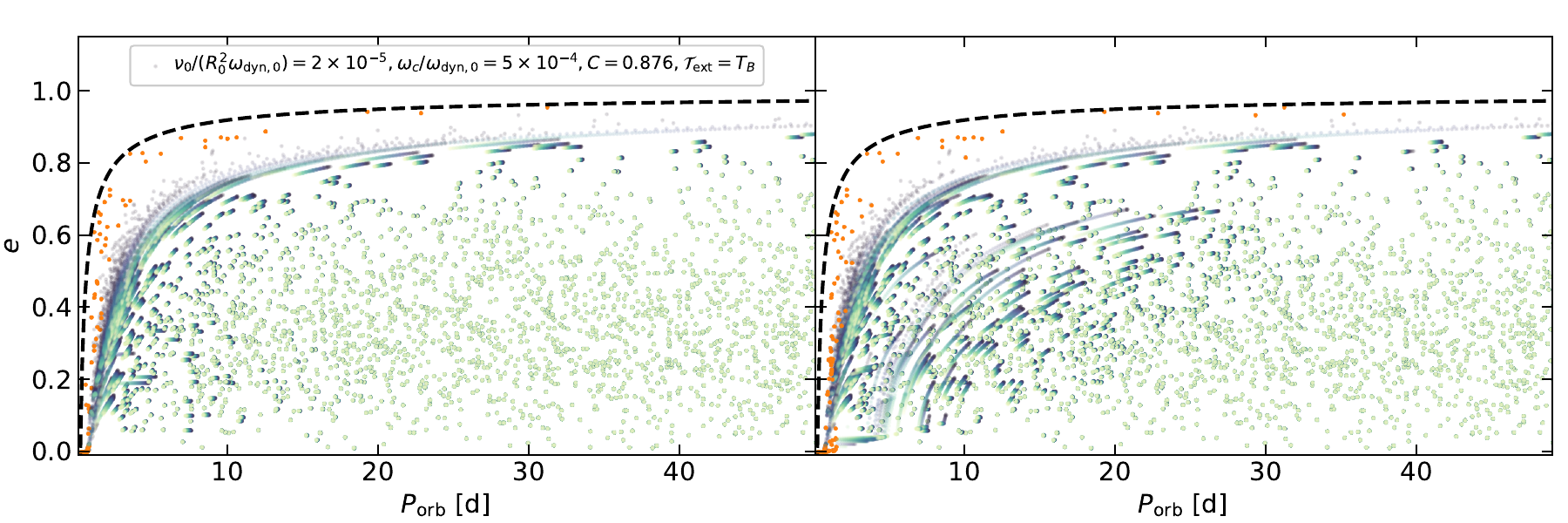}
    \includegraphics[width=\textwidth]{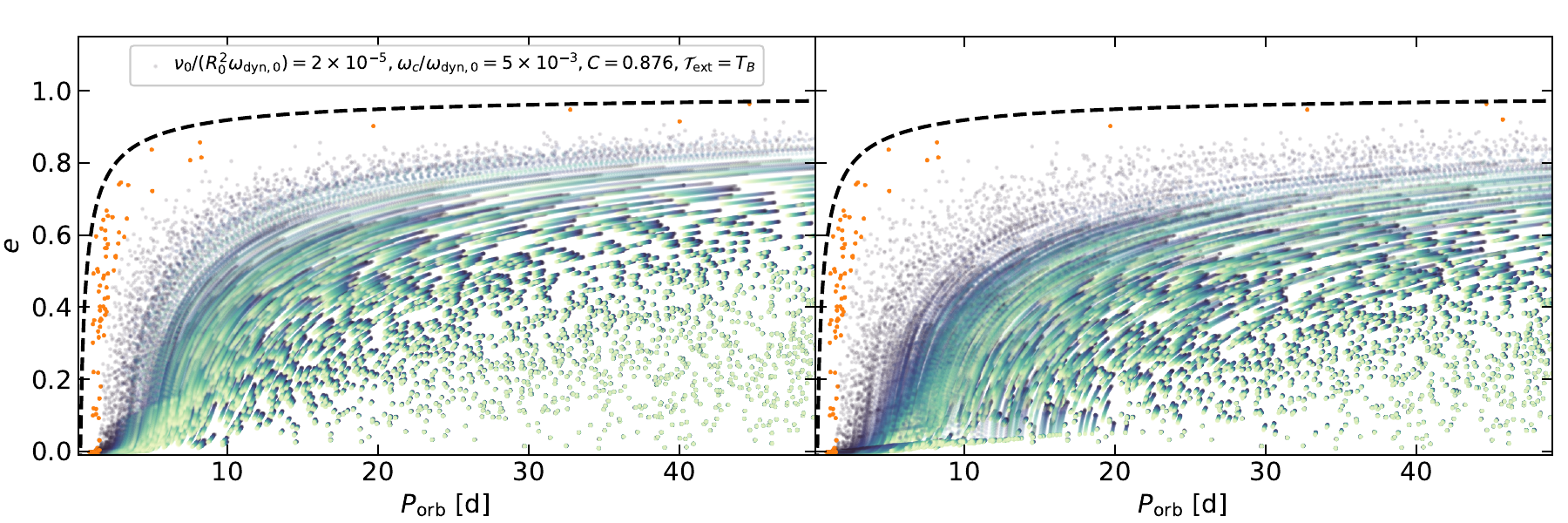}
    \caption{Eccentricities and orbital periods, sampled every $10$Myr from (polytropic model) simulations with initial orbital periods, eccentricities, and rotation periods drawn from distributions described in \autoref{sec:ps}. Point colors indicate runtime, from dark blue for $t=0$ to light green for $t=10^9$ years (the same opacity is used for every point). The black dashed line indicates eccentricities and periods where the pericentric radius $a(1-e)=R_0$, while orange points indicate simulations ended prematurely at the onset of common envelope evolution (in general $R\geq R_0$). The top panels show calculations without magnetic braking or high-frequency suppression, while both effects were included in generating the middle panels. The bottom panels show simulations run with a faster convective turnover time. The lefthand column shows calculations with a roughly equilibrium tidal theory, while the righthand column shows calculations including tidally driven inertial modes. Inertial modes enhance eccentricity damping, but suppression of the effective viscosity at large tidal frequencies affects the extent of the ``cool core'' of circular orbits produced by this damping.
    }
    \label{fig:ps_evPo}
\end{figure*}

The scatter plots shown in \autoref{fig:ps_evPo} plot values of eccentricity against orbital period for samples of simulations run without (top) and with (middle, bottom) the effects of both magnetic braking and high-frequency viscosity suppression. 

The top panels demonstrate that enhanced eccentricity damping by inertial modes (top right) can produce a ``cool core'' of nearly circular orbits that extends to larger orbital periods than in equivalent calculations with only f-mode tides (top left). In a large sample of main-sequence binaries reported by \emph{Gaia}, \citet{Dewberry2025} found that this cool core extends to $P_{\rm orb}\sim15-20$ days, which agrees reasonably well with that shown in \autoref{fig:ps_evPo} (top right). The middle panels in \autoref{fig:ps_evPo} demonstrate that high-frequency suppression of the eddy viscosity significantly slows this circularization. However, the bottom panels show that a cool core can be recovered by adopting a marginally faster convective turnover frequency.

\begin{figure*}[ht!]
    \centering
    \includegraphics[width=\textwidth]{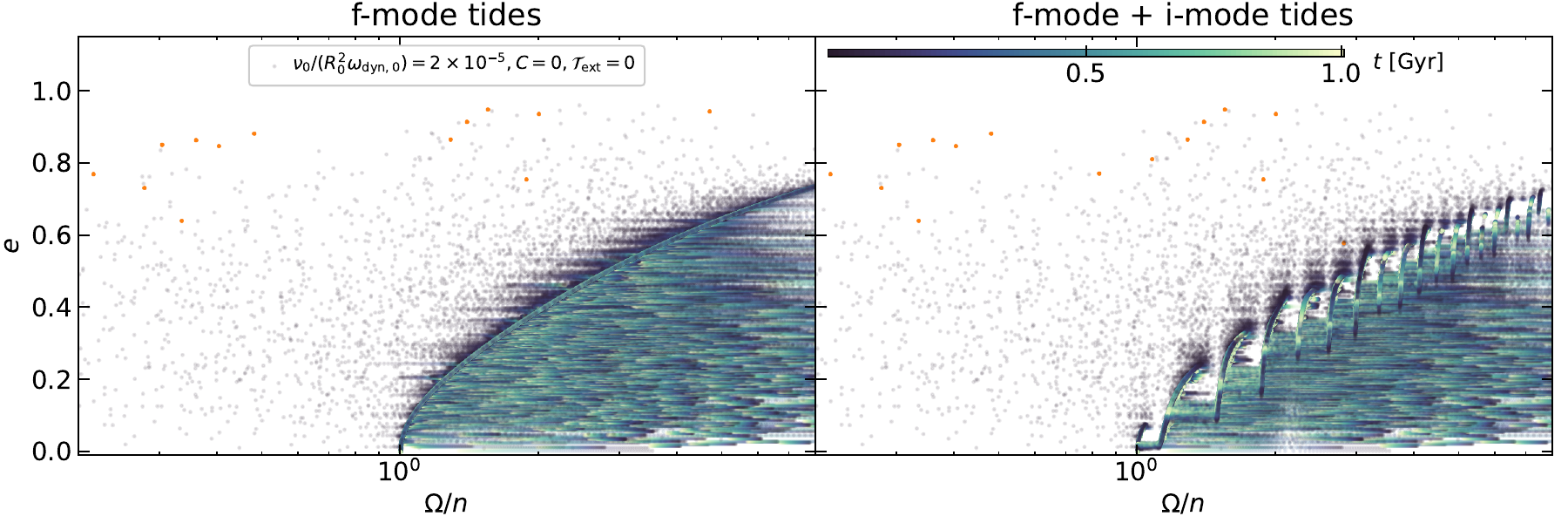}
    \includegraphics[width=\textwidth]{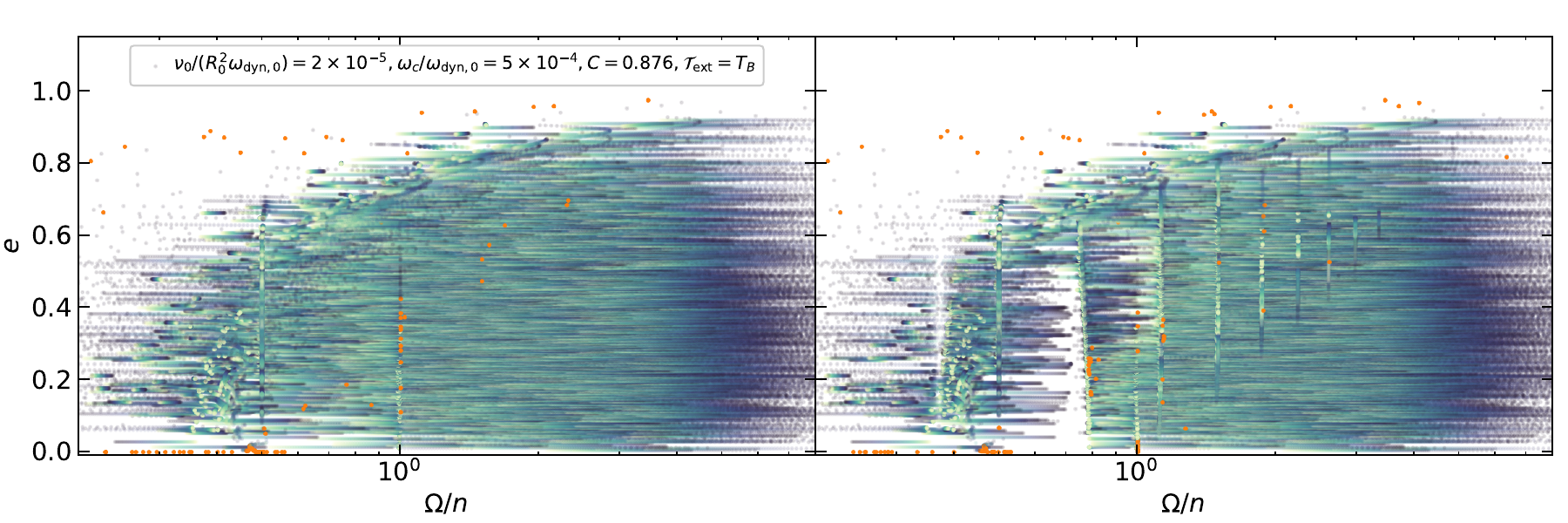}
    \includegraphics[width=\textwidth]{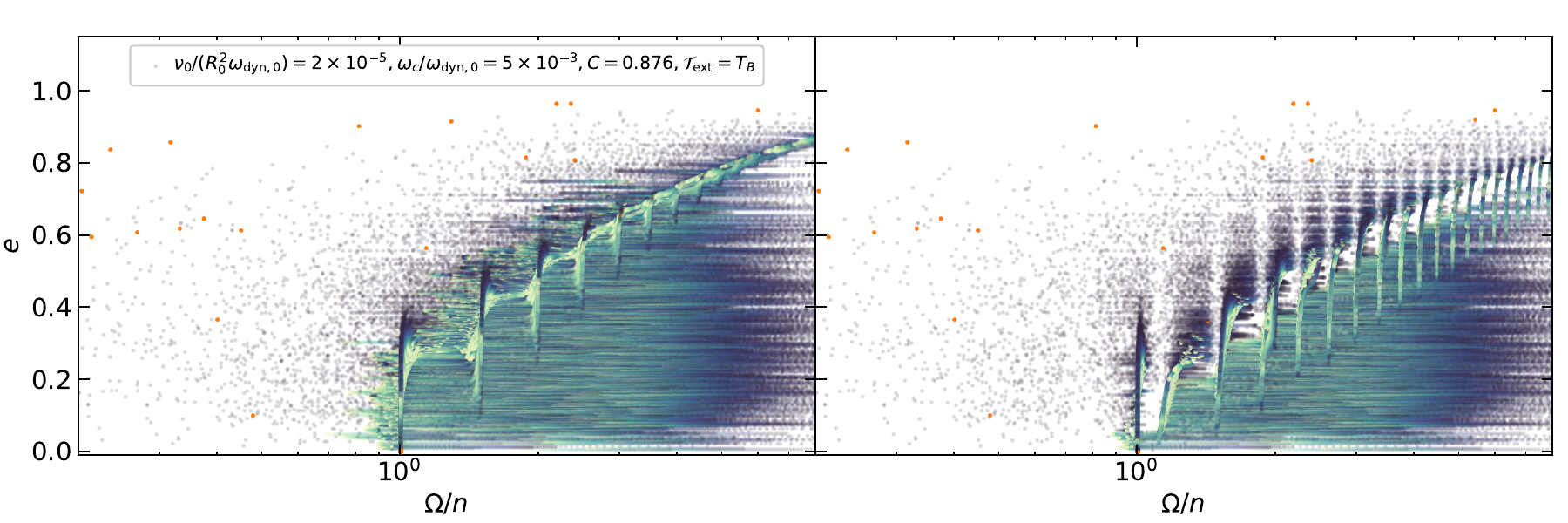}
    \caption{
    Eccentricities and ratios of the orbital period to rotation period, $P_{\rm orb}/P_{\rm rot}=\Omega/n$ for the same simulations shown in \autoref{fig:ps_evPo}. While simple f-mode tides (top left) lead to pseudosynchronization \citep{Hut1981}, both tidally driven inertial modes and high-frequency suppression produce vertical ridges of systems with discrete values of the period ratio. 
    }
    \label{fig:ps_evomn}
\end{figure*}

The lefthand panels of \autoref{fig:ps_evomn} are roughly consistent with the calculations of \citet{Barker2022}, who found that equilibrium tides cannot even circularize main-sequence binaries with orbital periods larger than a day, when eddy viscosities are suppressed at high frequencies (see their Fig. 3, bottom). On the other hand, \citet{Barker2022} predicted that tidally driven inertial waves could rapidly circularize all binaries with orbital periods less than $P_{\rm orb}\sim10$ days. This prediction appears to be somewhat more optimistic about the efficiency of circularization than the middle and bottom panels of \autoref{fig:ps_evomn}. Indeed, in \autoref{app:fav} we find that applying their analysis to the polytropic model considered here predicts circularization out to $P_{\rm orb}\approx17$ days, independent of the convective turbulence.

The discrepancy comes from our use of tidal frequency-dependent calculations, rather than the frequency-averaged response functions considered by \citet{Barker2022}. The frequency-averaged prediction is independent of the eddy viscosity because it derives from a model in which an impulsive force (from, e.g., a pericenter passage of a highly eccentric orbit) instantaneously transfers energy into a broad spectrum of inertial waves, irrespective of how they damp \citep{Ogilvie2013}. Our results demonstrate that the non-impulsive secular evolution does not follow immediately from frequency averaged quality factors, and can depend on the details of inertial wave dissipation (i.e., the eddy viscosity prescription). Then again, nonlinear evolution of the mean flow that deviates from the rigid rotation considered here may alter the dissipation predicted both by the frequency-averaged and frequency-dependent linear theory \citep{Astoul2022,Astoul2023,Astoul2025}.

Nevertheless, the righthand panels of \autoref{fig:ps_evPo} do show that frequency-dependent models of tidally driven inertial waves predict strongly enhanced circularization over equilibrium tides. Moreover, the comparison shown in \autoref{fig:inc_Ek6} suggests that the development of a central core should enhance inertial wave circularization further. Future work should aim to clarify the nonlinear interaction of convective turbulence with dynamical tides involving both inertial and gravito-inertial waves in more realistic stellar and planetary interior models.

\begin{figure*}[ht!]
    \centering
    \includegraphics[width=\textwidth]{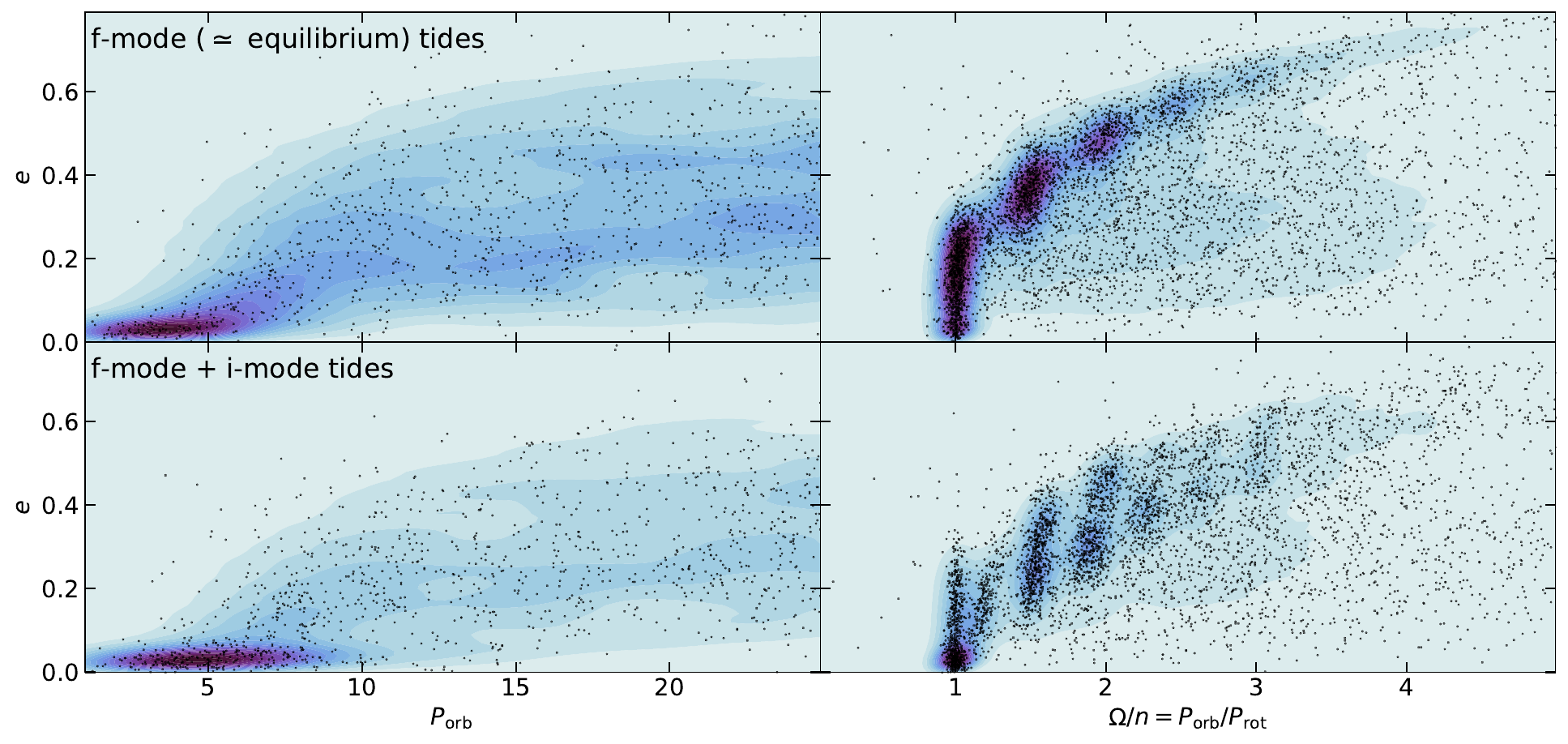}
    \caption{Mock populations of eccentricity, orbital period, and period ratio data constructed by aggregating and adding uncertainty to the simulation data shown in the bottom panels of \autoref{fig:ps_evPo} and \autoref{fig:ps_evomn}. Contours illustrate kernel density estimates, and the scatter plots show 5000 random draws from the larger data sample. The lefthand panels illustrate how tidally driven waves can produce an extended cool core of nearly circular binaries, while the righthand panels demonstrate discrete features they can produce in the period ratio $P_{\rm orb}/P_{\rm rot}$.
    }
    \label{fig:kdes}
\end{figure*}

\subsubsection{Asynchronization}
\autoref{fig:ps_evomn} demonstrates that tidally driven inertial waves might produce distinctive features in observed rotation rates. The panels in this figure plot eccentricity against the ratio of orbital to rotation periods, $P_{\rm orb}/P_{\rm rot}=\Omega/n$, for the simulations shown in \autoref{fig:ps_evPo}. While calculations including only equilibrium tides and a constant kinematic viscosity (top left panel) cluster around pseudosynchronization, those with tidally driven inertial waves and/or high-frequency suppression instead exhibit overdensities at discrete period ratios, which persist across broad ranges of eccentricity.

The exact values of these period ratios depend on the frequencies of the inertial waves of the tidally perturbed objects, which in our calculations are fixed by the fact that we perturb identical primary stars. In reality, these period ratios will vary with internal structure (and perhaps differential rotation), possibly ``smearing'' the narrowly concentrated vertical ridges shown in \autoref{fig:ps_evomn}. However, inertial wave frequencies are much more closely linked to rotation, and much less sensitive to other properties of internal structure than the frequencies of other oscillations (e.g., gravito-inertial modes). Their similarity across different stellar models with a given rotation profile may therefore lead to identifiable features in the observational data. 

Notably, the combination of both magnetic braking and tidally driven inertial modes produces populations of both (i) synchronous ($\Omega/n\simeq1$) rotators with relatively large, non-zero eccentricities ($e\lesssim 0.5$; see all of the middle and bottom panels in \autoref{fig:ps_evomn}), and (ii) sub-synchronous rotators (for example with the period ratio $\Omega/n\simeq0.8$; see the middle right panel). Populations of eccentric, synchronous rotators and sub-synchronous rotators have both been tentatively identified in observations of \emph{Kepler} and \emph{TESS} binaries \citep{Lurie2017,Hobson-Ritz2025,Yu2025}. These populations are difficult to explain with classical tidal theories, but can be naturally produced by the balance of torques considered here.

\subsubsection{Population-level predictions}
\autoref{fig:kdes} translates the simulations shown in \autoref{fig:ps_evPo} and \autoref{fig:ps_evomn} into rough predictions for population-level trends. The blue contours plot kernel density estimates of the probability density functions for eccentricity, orbital period, and period ratio, computed with aggregate data sampled every $10$Myr from the set of simulations run with high-frequency suppression, magnetic braking, and a faster convective turnover frequency $\omega_c/\omega_{\rm dyn,0}=5\times10^{-3}$. Before computing these kernel density estimates, we perturb orbital periods, rotation periods, and each component of eccentricity vectors with errors drawn from normal distributions. Somewhat optimistically, we set the standard deviations of these distributions to be $0.01P_{\rm orb}$, $0.03P_{\rm rot}$, and $0.025$ (respectively).

A wide range of simplifying assumptions underlie these synthetic data, not least of which being a non-evolving population of stars with identical masses and internal structures. However, \autoref{fig:kdes} highlights general trends that may still translate to observed binary populations. First of all, the lefthand panels clearly demonstrate inertial modes' enhancement to eccentricity damping, with a cool core extending to $\sim10$ days, vis-a-vis $\sim5$ days for purely f-mode ($\simeq$ equilibrium) tides. While this may not be enough to explain the cool core extending to $\sim15$ days exhibited by main-sequence \emph{Gaia} binaries \citep{Dewberry2025}, our calculations with a solid central core suggest that additional eccentricity damping in more realistic stellar models may make up the difference.

Secondly, the righthand panels in \autoref{fig:kdes}  reiterate that discrete features at fixed values of $P_{\rm orb}/P_{\rm rot}$ can be produced by both a frequency-dependent eddy viscosity (top) and tidally driven inertial modes (bottom). In more realistic populations of stars with different masses, ages, and internal structures, different properties of convection or different inertial mode frequencies might blur these discrete features. Inertial oscillations may be less susceptible to such blurring, however, since their frequencies are primarily set by a planet or star's rotation rate. 

\section{Summary and conclusions}\label{sec:conc}
We have developed and explored the consequences of simple but physical models for the coupled evolution of orbital eccentricity and rotation in stars and gaseous planets that are fully convective or have large convective envelopes. 

When the orbital angular momentum is large compared to the spin angular momentum of the tidally perturbed object, torques from tidally driven inertial waves cause rapid transitions between discrete ratios of the rotation rate $\Omega$ to the orbital mean motion $n$ (equivalently, the ratio of the orbital period to rotation period). This mechanism is similar to a resonance lock, which remains locked thanks to the scaling of inertial wave frequencies with the rotation rate of a nearly isentropic fluid \citep{Dewberry2024}. Here we have demonstrated through both analysis (\autoref{fig:edot_cpr}) and numerical experiments (Figures \ref{fig:n15_aoe_vs_t}-\ref{fig:n15_aoe_vs_t_mb}) that inertial waves' modification to spin evolution can enhance eccentricity damping by orders of magnitude over equilibrium tidal models employing the same underlying microscopic treatment of dissipation (in this case, an eddy viscosity attributed to convective turbulence). 

Suppression of the convective eddy viscosity by ``fast tides'' introduces additional equilibrium spin states (see Section \ref{sec:nu}) that reduce inertial waves' enhancement to eccentricity damping relative to weak frictional theories (\autoref{fig:n15_eaO_omc}). However, we find that an additional torque from magnetic braking can drive inertial waves to large enough amplitudes that they regain their advantage (\autoref{fig:n15_eaO_omc_mb}). These amplitudes nevertheless remain relatively small for the equal mass binaries considered in this paper (\autoref{fig:n15p_amps}). Nonlinear wave breaking may be more of a concern in exoplanets, since they are perturbed by host stars with much larger relative masses.

The majority of our calculations oversimplify the internal structure of the primary star by modeling it as a fully isentropic polytrope. The presence of a solid or stably stratified core can considerably complicate the frequency dependence of the inertial wave tidal response (see \autoref{fig:demo_iks}, bottom). Despite this, we find that because inertial wave frequencies scale with the changing rotation rate, this complicated frequency dependence hastens but does not significantly alter the spin evolution and eccentricity damping exhibited by our polytropic models (\autoref{fig:inc_Ek6}). 

In an effort to identify distinctive observational signatures of tidal evolution, we have run an ensemble of simulations starting from observationally motivated initial values for eccentricity, orbital period, and rotation rate. These simulations demonstrate first that inertial waves' enhancement to eccentricity damping can produce an enhanced ``cold core'' of nearly circular systems (\autoref{fig:ps_evPo} and \autoref{fig:kdes}, left). This enhancement may help to reconcile theory with observations of enhanced circularization out to orbital periods of $\sim10-15$ days in main-sequence, solar-type binaries. Enhanced eccentricity damping by tidally driven inertial waves may also be important for the migration of hot and warm Jupiters.

Separately, the tidal mechanisms highlighted in this paper make independent observational predictions for the ratio of orbital to rotation period, $P_{\rm orb}/P_{\rm rot}=\Omega/n$. Classical tidal models \citep[e.g., appealing to constant time lag][]{Hut1981} predict pseudosynchronization at a period ratio that increases monotonically with eccentricity (red curve in the middle panels of Figures \ref{fig:n15_eaO}-\ref{fig:n15_eaO_omc_mb}). In contrast, we find that inertial wave torques, as well as high-frequency suppression of the eddy viscosity, lead to pileups at discrete period ratios that are \textit{independent of eccentricity} (see the vertical ridges in \autoref{fig:ps_evomn} and \autoref{fig:kdes}, right). These features could help to explain observed populations of sub-synchronously rotating, as well as synchronously rotating but still eccentric binaries.

\begin{acknowledgments}
I thank the anonymous referee for an incisive and helpful review. I also thank Yanqin Wu, Jim Fuller, and Adrian Barker for very helpful conversations. 
\end{acknowledgments}

\software{
    \texttt{numpy} \citep{numpy}, 
    \texttt{matplotlib} \citep{matplotlib}, 
    \texttt{scipy} \citep{scipy}, 
    \texttt{MESA} \citep{Paxton2011, Paxton2013, Paxton2015, Paxton2018, Paxton2019, Jermyn2023},
    \texttt{REBOUND} \citep{rebound}
}


\appendix

\section{Hansen series}\label{app:hansum}
Defining 
\begin{equation}
    X_{mk}(e)
    =\frac{1}{\pi}
    \int_0^{\pi}(1 - e\cos E)^{-2}
    \cos\left[ 
        2m\tan^{-1}\left(\sqrt{\frac{1+e}{1-e}}\tan \frac{E}{2}\right)
        -k(E-e\sin E)
    \right]
    \text{d}E,
\end{equation}
the series $H_{pm}=\sum_kk^p|X_{mk}|^2$ can be found as 
\citep[e.g.,][]{Leconte2010,Burkart2012}
\begin{align}
    H_{0m}(e)
    &=\sum_{k=-\infty}^\infty |X_{mk}|^2
    =(1 - e^2)^{-9/2}\left[1 + 3e^2 + (3/8)e^4\right],
\\
    H_{12}(e)
    &=\sum_{k=-\infty}^\infty k|X_{2k}|^2
    =2(1 - e^2)^{-6}\left[1 + (15/2)e^2 + (45/8)e^4  + (5/16)e^6\right],
\\
    H_{20}(e)
    &=\sum_{k=-\infty}^\infty k^2|X_{0k}|^2
    =(1-e^2)^{-15/2}
    \left[
        (9/2) e^2 
        +(135/8) e^4 
        +(135/16) e^6 
        +(45/128) e^8
    \right],
\\
    H_{22}(e)
    &=\sum_{k=-\infty}^\infty k^2|X_{2k}|^2
    =(1-e^2)^{-15/2}
    \left[
        4
        +(121/2) e^2 
        +(975/8) e^4 
        +(695/16) e^6 
        +(185/128) e^8
    \right].
\end{align}

\section{Oscillation mode expansions}\label{app:modes}
\begin{figure*}
    \centering
    \includegraphics[width=\textwidth]{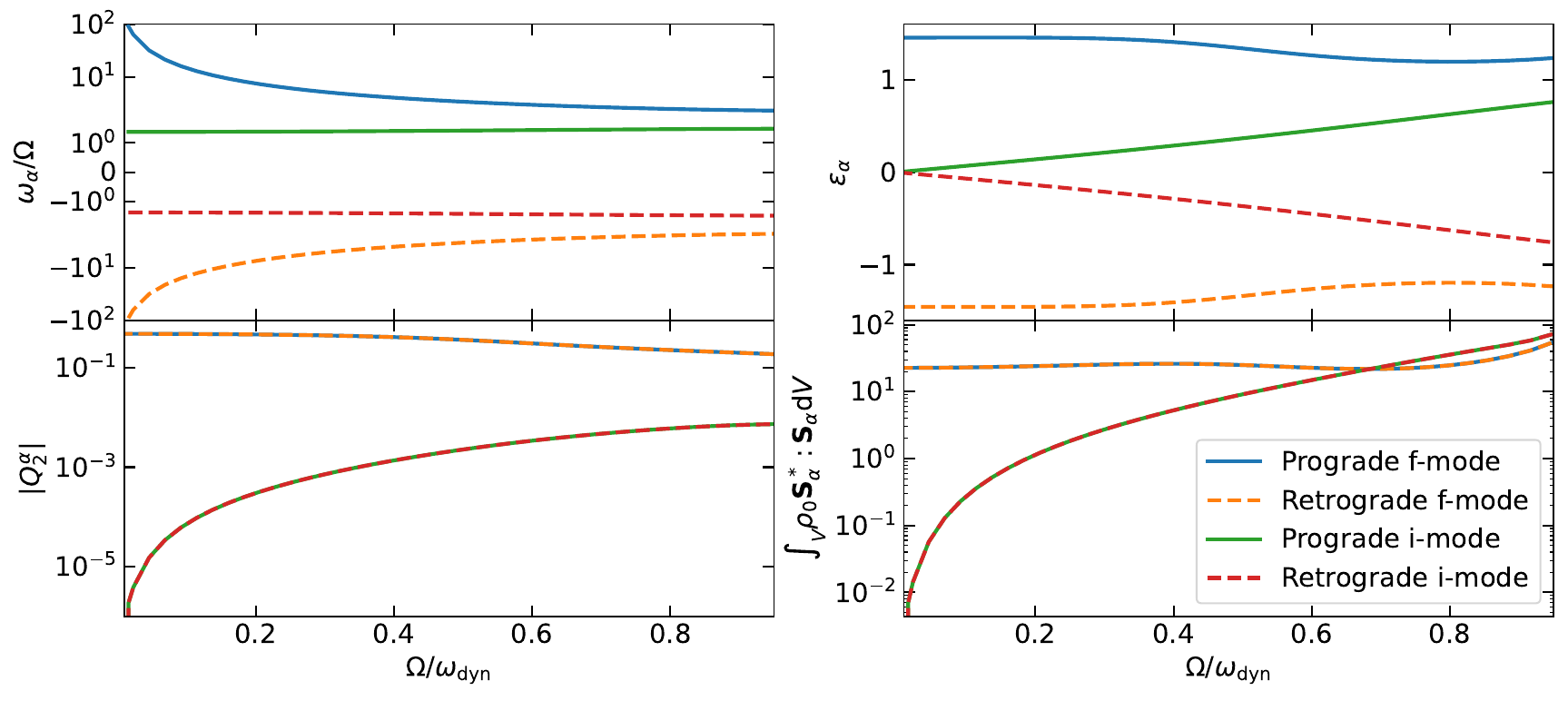}
    \caption{Analogue to Fig. 8 in \citetalias{Dewberry2024}, for $m=0$ oscillation modes (see Appendix C in that paper for definitions).}
    \label{fig:m0mode}
\end{figure*}

\autoref{fig:m0mode} plots the properties of $m=0$ f-modes and inertial modes required \citep[in addition to the $m=2$ mode properties presented in ][]{Dewberry2024} to compute the tidal power.

\section{Frequency-averaged estimate}\label{app:fav}
An impulsive model of low-frequency inertial wave excitation in an isentropic $n=1.5$ polytrope gives the frequency average 
\begin{equation}
    I_{22}
    \equiv
    \left(\frac{\omega_{\rm dyn}}{\Omega}\right)^2
    \int_{-\infty}^{\infty} {\rm Im}[k_{22}]
    \frac{{\rm d}\omega}{\omega}
    \simeq 0.011,
\end{equation}
ignoring the effects of centrifugal flattening \citep{Ogilvie2013}. For the non-evolving stellar models with $M=M'=M_\odot$ and $R=2R_\odot$ considered in \autoref{sec:ps}, the analysis of \citet{Barker2022} predicts that in spin-synchronized systems the eccentricity will evolve according to
\begin{equation}
    \ln e(t_0)-\ln e(t_1)
    \approx
    \frac{225\pi}{8}
    q\left(\frac{1}{1 + q}\right)^{5/3}
    P_{\rm dyn}^{16/3}P_{\rm orb}^{-19/3}I_{22}
    \int_{t_0}^{t_1}{\rm d}t
    \approx0.32 
    P_{\rm dyn}^{16/3}P_{\rm orb}^{-19/3}
    (t_1 - t_0),
\end{equation}
where $t_0$ and $t_1$ are initial and final times, and $P_{\rm dyn}=2\pi/\omega_{\rm dyn}.$ Setting $t_0=0$, $t_1=1$Gyr, $e(t_0)=1,$ and $e(t_1)=0.01$, this equation predicts circularization out to orbital periods of $P_{\rm orb}\simeq17.16$ days.


\bibliography{iwecc}{}
\bibliographystyle{aasjournal}



\end{document}